\documentclass[11pt,oneside,openleft]{book}

\usepackage[letterpaper,lmargin=2.9cm,rmargin=1.9cm,bmargin=2.8cm,tmargin=2.8cm,includeheadfoot]{geometry} 

\usepackage{graphicx}  
\usepackage{tikz, tikz-3dplot}
\usepackage{flafter}  
\usepackage[list=off,font=small,skip=3pt]{caption}
\usepackage{subcaption}

\usepackage{lipsum}

\usepackage{cite}

\usepackage{amsthm}
\usepackage{amsmath,amssymb}  
\usepackage{textcomp} 
\usepackage{bm}  

\usepackage[utf8]{inputenc} 
\usepackage[T1]{fontenc}
\usepackage[spanish,slovene]{babel}

\usepackage{booktabs}
\usepackage{enumitem}
\setlist{itemsep=0.5pt}

\usepackage[plain]{fancyref}
\usepackage[bookmarksopen,colorlinks,pdftex]{hyperref}  
\hypersetup{citecolor=black, filecolor=black, linkcolor=black, urlcolor=black} 

\usepackage{fancyhdr}
\usepackage{aas_macros} 
\usepackage[]{tocbibind} 

\makeatletter 
\DeclareRobustCommand*{\bfseries}{%
  \not@math@alphabet\bfseries\mathbf
  \fontseries\bfdefault\selectfont
  \boldmath
}
\makeatother

\begin{document}
	
	\selectlanguage{spanish}

    \renewcommand{\tablename}{Tabla}

    \renewcommand{\thefootnote}{\Roman{footnote}}

    \lefthyphenmin=2
    \righthyphenmin=2

    \frontmatter

    \pagestyle{empty}

\newcommand{\HRule}{\rule{\linewidth}{1pt}} 

\begin{titlepage}
	
				
\begin{center}

\vfill

\Large

Universidad de La Habana

\medskip

Facultad de Física

\vspace{10mm}

\centerline{\mbox{
\includegraphics[height=50mm,keepaspectratio]{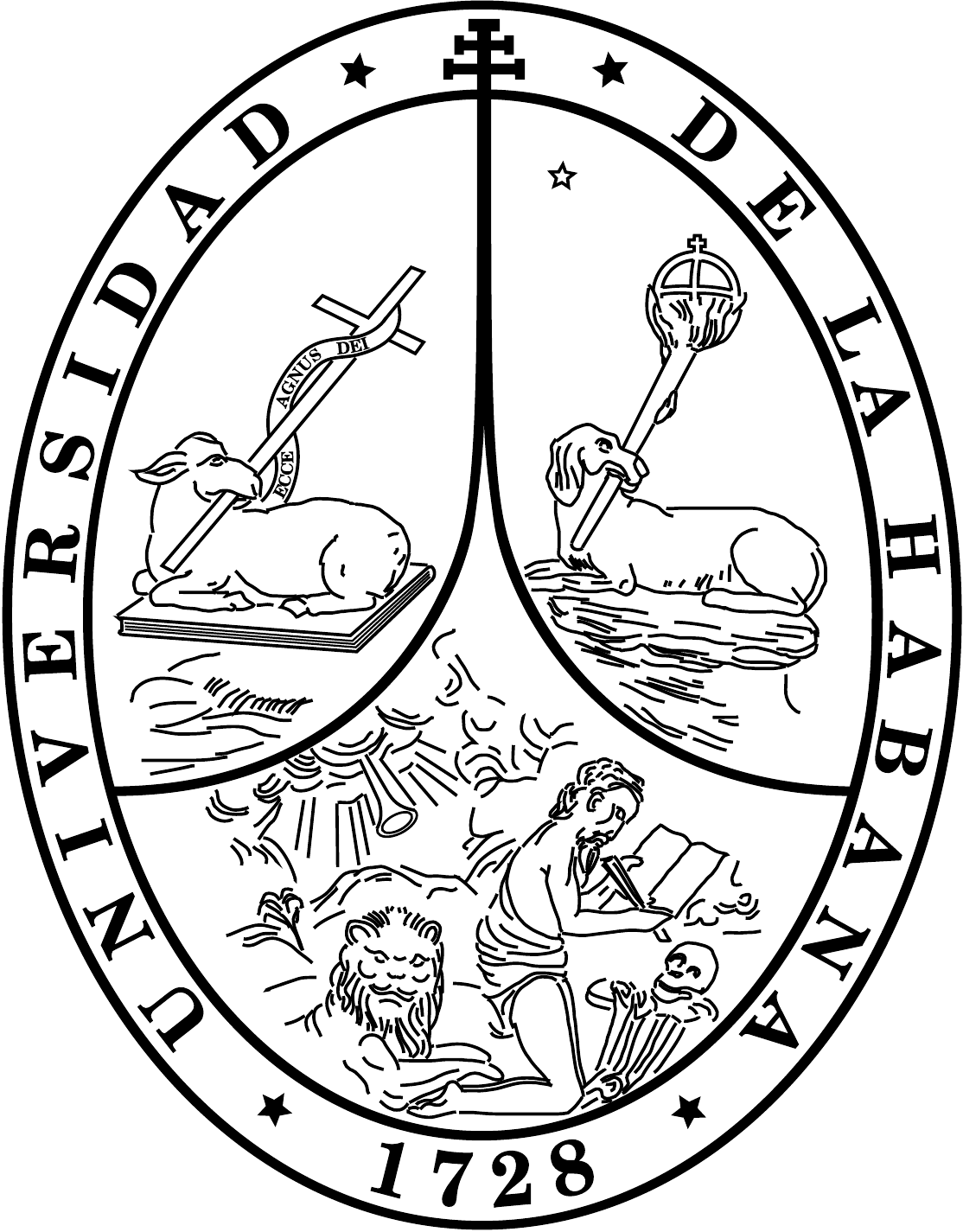}}}

\vfill

{\Large TESIS  \\[5pt] \emph{presentada en opción al grado científico de}  \\[5pt] \textbf{Máster en Ciencias Físicas}} \\[20pt]


{\LARGE \bfseries \MakeUppercase{Termodinámica de un gas magnetizado de bosones vectoriales neutros} \\[5pt] \MakeUppercase{}}




\vfill

\begin{tabular}{rl}

\textbf{Autora:} & Lic. Lismary de la Caridad Suárez González \\
\noalign{\vspace{10pt}}

\textbf{Tutoras:} & Dra. Gretel Quintero Angulo, FF-UH \\[5pt]
 & Dra. Aurora Pérez Martínez, ICIMAF \\[5pt]
			      
\noalign{\vspace{2mm}}
\end{tabular}

\vfill

\Large
La Habana, 2020

\end{center}


\end{titlepage}
\cleardoublepage

   	\begin{center}
	\section*{Agradecimientos}
\end{center}


\begin{quote}

\begin{flushright}
Mis primeras l\'ineas de agradecimiento tienen que ser para mis tutoras porque sencillamente son las mejores. Primeramente, a Gretel, por proponerme la genial idea de trabajar en temas de astrof\'isica, y a Aurora, por aceptarme como estudiante de maestría aun sabiendo que no ten\'ia ninguna formación teórica.	Por el conocimiento compartido y sobre todo por el tiempo dedicado. Por la paciencia en mis per\'iodos de vagancia. A Gretel, en particular, por la paciencia infinita con mis errores tipográficos que por cierto también parecían infinitos. Por la amistad, y por los momentos de diversión;
\end{flushright} 

\begin{flushright}
	A mis padres, por el amor, el apoyo, el \'animo y por la confianza depositada en m\'i;
\end{flushright}

\begin{flushright}
	Al ICIMAF, por abrirme las puertas y brindarme un lugar donde continuar con mi vocación científica. En general, por todas las oportunidades que han contribuido a mi formaci\'on profesional;  
\end{flushright}

\begin{flushright}
	A mis compa\~{n}eros del departamento de Física Teórica del ICIMAF: Diana, Yamila, Gaby, Duvier, Dariel, Samantha, Elizabeth, Jorgito, Hugo, Augusto y Cabo, porque es un grupo de trabajo donde el compa\~{n}erismo sobra y es muy agradable trabajar en ese tipo de ambiente. En especial quiero agradecer a Diana quien fue de mucha ayuda en mis inicios en LaTex y en Mathematica;
\end{flushright} 

\begin{flushright}
	A Mercy, secretaria de la Facultad de Física, por la dulzura y amabilidad mostrada en lo referente a todos los temas burocráticos de la maestría;
\end{flushright}

\begin{flushright}
 A Sergio, por el amor y cari\~{n}o en las etapas finales de la tesis. A Gabriela, por ser mi mejor amiga y aguantarme en mis periodos de estrés y, sobre todo, por estar siempre para m\'i cuando necesito hablar y desahogarme. A Rey, por prestarme sus computadoras para realizar mis cálculos m\'as rápido ;
\end{flushright}

\begin{flushright}
	A todos mi amigos del grupo de teatro "Monopolo Magnetico'': Gretel, Joeluis, Sandra, Vicente, Adri\'an, Alejandro, Landy y Yanela, y de mi banda "Jones Rock Band'': Tony, Rey Luis, Ernesto, Javi y B\'arbaro. Porque todos ellos hacen que mi vida sea m\'as alegre y bonita.
\end{flushright}

\end{quote}

\vfill
\cleardoublepage

\pagestyle{plain}

    \pagestyle{empty}

\vfill

\begin{center}
\section*{Resumen}
\addcontentsline{toc}{chapter}{Resumen}
\end{center}
\medskip

\begin{quote}

Esta tesis está dedicada al estudio de las propiedades termodinámicas de un gas magnetizado de bosones vectoriales neutros a toda temperatura, con el fin de proporcionar ecuaciones de estado que permitan descripciones más generales y precisas de los fenómenos astrofísicos. Para ello, a partir del espectro energético derivado de la teoría de Proca, se obtienen expresiones analíticas para las magnitudes termodinámicas válidas a toda temperatura, así como sus límites no relativistas. A través de estas expresiones, y considerando el sistema en condiciones astrofísicas (densidades de partículas, temperaturas y campos magnéticos en el orden de los posibles en las Estrellas de Neutrones), se investigan la condensación de Bose-Einstein, las propiedades magnéticas y las ecuaciones de estado del gas, haciendo especial énfasis en la influencia de las antipartículas y el campo magnético. En todos los casos los resultados se comparan con sus análogos obtenidos en los límites de baja temperatura y no relativista, pues ello permite establecer los rangos aproximados de validez de dichas aproximaciones y lograr una mejor comprensión de sus efectos en el sistema estudiado.

\end{quote}


\begin{center}
\section*{Abstract}
\addcontentsline{toc}{chapter}{Abstract}
\end{center}
\medskip

\begin{quote}

This thesis is dedicated to study the thermodynamic properties of a magnetized neutral vector boson gas at any temperature, with the aim to provide equations of state that allow more general and precise descriptions of astrophysical phenomena. The all--temperature analytical expressions  for the thermodynamic magnitudes, as well as their non-relativistic limits, are obtained starting from the energy spectrum given by Proca's theory. With these expressions, and considering the system under astrophysical conditions (particle densities, temperatures and magnetic fields in the order of the estimated for Neutron Stars), we investigate the Bose-Einstein condensation, the magnetic properties and the equations of state of the gas, making a special emphasis on the influence of antiparticles and magnetic field. In all cases, the results are compared with their analogues in the low temperature and the non-relativistic limits. This allows us to establish the ranges of validity of these approximations and to achieve a better understanding of their effects on the studied system.

\end{quote}

\vfill
\cleardoublepage

    \tableofcontents


    \mainmatter

    \pagestyle{fancy}
    \pagestyle{plain}

\chapter*{Introducción}
\label{intro}
\addcontentsline{toc}{chapter}{Introducción}

Uno de los problemas más desafiantes de la física moderna consiste en el estudio de la materia en condiciones extremas -densidades supranucleares y campos magnéticos fuertes-, y la determinación de las ecuaciones de estado (EdE) asociadas a ella \cite{weber2017pulsars}. La interacción fuerte, que es la dominante a densidades nucleares o superiores, se explica actualmente a través de la cromodinámica cuántica (QCD por sus siglas en inglés de \textit{Quantum Chromodynamics}). Esta teoría tiene la limitante de que no permite usar m\'etodos perturbativos para la descripci\'on de hadrones con masas menores que $2$ GeV \cite{buballa2005njl}.
Desde el punto de vista experimental, el problema tambi\'en es complejo, debido a que todavía no se ha logrado obtener materia a densidades mayores que la densidad de saturación nuclear $\rho_N\simeq2.4\times10^{14}g/cm^3$. Por otra parte, se sabe que existen configuraciones estelares estables que contienen materia en una de las formas m\'as densas que se puede encontrar en el Universo. De modo que, hoy en día, los entornos astrofísicos son unos de los mejores escenarios para investigar las propiedades de la materia superdensa. En particular, las estrellas de neutrones (ENs) son excelentes laboratorios naturales, pues son objetos con tiempo de vida prácticamente infinito, cuyas densidades pueden llegar a ser de hasta un orden de magnitud mayor que las de los núcleos atómicos\cite{weber2017pulsars}.

La idea de una estrella compuesta esencialmente por neutrones fue desarrollada en 1934 por los astrónomos Baade y Zwicky, a solo dos a\~{n}os del descubrimiento del neutr\'on \cite{weber2017pulsars}. Siguiendo esa propuesta, Tolman, Oppenheimer y Volkoff llevaron a cabo los primeros cálculos teóricos de la estructura macrosc\'opica de estas estrellas \cite{oppenheimer1939massive}. Sin embargo, tuvieron que pasar más de 30 años hasta que, en 1967, la estudiante de doctorado Jocelyn Bell, analizando observaciones en ondas de radio, descubriera en el cielo una serie de pulsos coherentes con un período corto y muy regular, asociados a una fuente puntual, a la que se denominó “pulsar”(\textit{pulsating star}) y que apenas un año después fue identificada como una estrella de neutrones altamente magnetizada en rotaci\'on \cite{weber2017pulsars}. En la actualidad, más de 1500 púlsares han sido detectados solamente en nuestra galaxia \cite{Camenzind}.

Las ENs tienen masas del orden de $M\sim1.5 M_{\odot}$\footnote{$M_{\odot}=1.989\times10^{30}$kg es la masa del Sol.}, radios de $R\sim10$ km, densidad de masa bari\'onica $\rho\sim10^7-10^{15}g/cm^{3}$, temperaturas $T\sim10^5-10^{11}$ K, y campos magnéticos que alcanzan valores entre $10^9-10^{15}$G en su superficie y hasta $10^{18}$G en su interior \cite{lattimer2007neutron}. Su estructura interna se divide en capas. De afuera hacia adentro encontramos \textit{la atmósfera} de unos centímetros de espesor; \textit{la envoltura} que es una capa de alrededor de unos cientos de metros, con densidades que varían entre $10^4g/cm^{3}<\rho<10^6g/cm^{3}$ y materia compuesta por n\'ucleos atómicos y electrones no relativistas; \textit{la corteza}, con aproximadamente 1km de grosor, que se separa en exterior e interior; y el núcleo. En la \textit{corteza externa}, a densidades tales que $ 7\times 10^6 g/cm^3<\rho<4.3\times 10^{11} g/cm^3$, los electrones se vuelven relativistas, mientras que los n\'ucleos at\'omicos (metales m\'as ligeros), forman una red s\'olida. En la \textit{corteza interna}, las densidades oscilan entre $4.3\times10^{11} g/cm^3$ y $2\times10^{14}g/cm^3$ aproximadamente y los electrones comienzan a penetrar en el n\'ucleo at\'omico produciendo el decaimiento-$\beta$ inverso. A medida que el número de neutrones aumenta en los núcleos atómicos, estos se vuelven más densos y llega un punto ($\rho\sim4.3\times10^{11} g/cm^3$) en el que los neutrones comienzan a brotar fuera el núcleo en lo que se conoce como el goteo de neutrones \cite{Camenzind}. El núcleo de la estrella ocupa el $90\%$ de su volumen y contiene la mayoría de su masa, pudiendo alcanzarse en él densidades superiores a $\rho_N$. A pesar de que las ENs han sido ampliamente estudiadas todavía no existe un consenso en cuanto a la materia que forma su núcleo. Esto se debe a la imposibilidad de realizar experimentos a densidades supranucleares que descarten o corroboren los diferentes modelos te\'oricos propuestos, y a que las observaciones aún no son concluyentes. 

El punto de partida en la descripci\'on te\'orica de las ENs es siempre el estudio termodinámico de la materia que la compone, ya que a partir de este se obtienen las ecuaciones de estado  que se utilizan en el cálculo de sus observables macroscópicos (masa, radio, campo magnético, momento de inercia, momento cuadrupolar de masa, período de rotación, etc.). Por ello resulta en extremo importante que dichos estudios sean lo más realistas y completos posible. Todo modelo teórico de las ENs parte de suponer el n\'ucleo compuesto por un gas de neutrones, protones y electrones con una peque\~{n}a fracci\'on de muones \cite{Camenzind}. Sin embargo, a densidades mayores que la densidad nuclear, los momentos de Fermi de los nucleones son tan altos que pueden ocurrir reacciones de creación de otras part\'iculas como hiperones o mesones\cite{weber2017pulsars,baldo2003neutron}, o surgir fases ex\'oticas de la materia como el plasma de quarks y gluones, el superfluido de nucleones o las fases superconductoras de color  \cite{weber2017pulsars,baldo2003neutron,chavanis2012bose,quintero2017anisotropic,latifah2014bosons}. 

Una de las hip\'otesis m\'as populares en la modelaci\'on del n\'ucleo de las ENs, es aquella que considera que en \'el los protones y neutrones se encuentran apareados \cite{chavanis2012bose,latifah2014bosons,quintero2019self}. En dependencia de la fortaleza de atracción entre los fermiones, los pares pueden comportarse como pares de Cooper (fermiones débilmente ligados) o bosones efectivos (fermiones fuertemente ligados) \cite{sedrakian2006nuclear,gusakov2004enhanced,lombardo2005superfluid}. Aunque los modelos derivados de estos dos casos, a saber, los de estrellas de neutrones con interior superfluido y los de estrellas de bosones, datan de más de cincuenta años, ellos han vuelto a tomar auge en la última década \cite{chavanis2012bose,latifah2014bosons,chamel2017superfluidity,sedrakian2019superfluidity,pethick2015bose}. Esto último gracias a los excelentes resultados obtenidos al ajustar la curva de enfriamiento del objeto compacto que se encuentra en el centro de Cassiopea A con un modelo de EN con interior superfluido  \cite{shternin2011cooling}, y a la demostración experimental de que la superfluidez y la condensación de Bose-Eisntein (CBE) son los estados extremos del fenómeno de apareamiento de fermiones \cite{sun2010relativistic,sedrakian2006nuclear,astrakharchik2005momentum,leggett2012bec}. Recientemente, en el grupo de investigaci\'on al que se adscribe esta tesis, varios estudios han sido dedicados a este tipo de modelo para el n\'ucleo de las ENs formadas por neutrones apareados que se comportan como bosones efectivos \cite{angulo2017thermodynamic,quintero2017anisotropic,quintero2019self}. En particular, referidos a los efectos de los altos campos magn\'eticos de estas estrellas en sus propiedades macrosc\'opicas y microsc\'opicas, y a la b\'usqueda de mecanismos que expliquen la generaci\'on de los mismos.   

Incluir al campo magn\'etico en la modelaci\'on de las ENs es un paso crucial en el camino hacia modelos cada vez m\'as realistas, pues estas estrellas son objetos fuertemente magnetizados, y se ha demostrado que las propiedades magnéticas de las partículas que las componen tienen una influencia importante en su fenomenología y estructura \cite{alvear2019anisotropic,terrero2019modeling,perez2019modeling,quintero2019self}. En el caso de las estrellas de bosones estudiadas en \cite{tesisgretel} el campo magn\'etico cobra gran importancia porque a las densidades t\'ipicas del núcleo de estos objetos, los pares de neutrones que se forman son vectoriales \cite{sedrakian2019superfluidity,baldo19983,tamagaki1970superfluid}.

Por otra parte, la presencia del campo magn\'etico enriquece much\'isimo la fenomenolog\'ia de los gases bos\'onicos de spin uno. En el caso de bosones vectoriales cargados, la cuantizaci\'on del momento perpendicular de las part\'iculas en niveles de Landau induce un cambio en la naturaleza de la transici\'on de fase al condensado, que deviene difusa, es decir, no ocurre  a una temperatura crítica definida sino que lo hace gradualmente en un intervalo de temperatura \cite{ROJAS1996148, Khalilov1997,Khalilov1999,PEREZROJAS2000,Rojas1996BoseEinsteinCM}. Por el contrario, en el caso de gases magnetizados de bosones vectoriales neutros la transición de fase al condensado es normal y se ve favorecida por la presencia del campo magn\'etico \cite{tesisgretel,angulo2017thermodynamic}. Pero ya sea el gas cargado o neutro, los gases bos\'onicos de spin uno presentan una interesante propiedad conocida como ferromagnetismo de Bose-Einstein, que consiste en la aparición de una magnetizaci\'on espontánea a $B=0$, un fenómeno que podría estar conectado con el origen de los campos magnéticos estelares \cite{angulo2017thermodynamic,yamada1982thermal,Rojas1996BoseEinsteinCM,PEREZROJAS2000}.

En los trabajos previos sobre el gas magnetizado de bosones vectoriales neutros\footnote{Bosones vectoriales neutros pueden ser mesones, átomos, y otros pares de fermiones con carga neta cero
y spin uno.} y los modelos de estrellas asociados a ellos, llevados a cabo en nuestro grupo de investigaci\'on, la descripci\'on termodinámica de este sistema de part\'iculas fue hecha en el l\'imite de baja temperatura ($T<<m$). Como veremos en la tesis, el límite de baja temperatura es equivalente a hacer una aproximación de campo magnético fuerte, que al aplicarse a pares de bosones compuestos por dos neutrones requerir\'ia el uso de campos a partir de $B\sim 10^{19}$G, un orden por encima que los campos magn\'eticos m\'as intensos esperados en el interior de las ENs. 
De ah\'i que extender a toda temperatura los estudios termodinámicos previos del gas de bosones vectoriales neutros (GBVN) \cite{tesisgretel} implicar\'ia  adem\'as obtener expresiones exactas en la regi\'on de campo magn\'etico d\'ebil. Este fue el propósito inicial de la tesis, sin embargo, a medida que avanzamos en los cálculos nos dimos cuenta de que ir más allá del límite $T<<m$ no solo permite obtener expresiones más generales y realistas para las EdE en esa regi\'on de temperatura, sino también estudiar la f\'isica de la regi\'on de alta temperatura, y en especial, la contribución de las antipartículas, usualmente despreciadas. Si bien es cierto que en el caso de bosones compuestos por dos neutrones temperaturas tales que $T\gtrsim m$ ($T\gtrsim 10^{13}$K) no son realistas (son demasiado altas tanto para los entornos astrof\'isicos \cite{Camenzind} como para existencia del par\cite{sedrakian2019superfluidity}), la caracterización del GBVN a toda temperatura s\'i podr\'ia ser relevante en el caso de bosones m\'as ligeros, y ser útil en otras ramas de la física, como la física de materia condensada \cite{Simkin_1999}, y la física de colisionadores de iones pesados \cite{ayala1997density,begun2006particle,su2008thermodynamic}. De modo que el estudio realizado en la tesis rebas\'o nuestras pretensiones originales y demostr\'o ser interesante en s\'i mismo.

De acuerdo con todo lo anterior, el \textbf{objetivo general de la presente tesis es estudiar las propiedades termodinámicas de un gas magnetizado de bosones vectoriales neutros a toda temperatura}, a fin de proporcionar ecuaciones de estado que permitan descripciones más generales y precisas de los objetos y fenómenos astrofísicos. Para ello nos proponemos los siguientes objetivos específicos:

\begin{itemize}
	\item Obtener una expresión para toda temperatura del potencial termodinámico de un gas magnetizado de bosones vectoriales neutros en el límite no relativista.
	
	\item Obtener las ecuaciones de estado de este gas y estudiar sus propiedades termodinámicas.

	\item A partir de la expresión obtenida en \cite{tesisgretel} para toda temperatura del potencial termodinámico de un gas magnetizado de bosones vectoriales neutros relativistas, obtener las ecuaciones de estado de dicho gas a toda temperatura y estudiar sus propiedades termodinámicas.
	
	\item Comparar los resultados obtenidos para el gas relativista de bosones vectoriales neutros a toda temperatura, con los resultados en los límites no relativista y de baja temperatura para determinar los rangos de validez de dichas aproximaciones y sus efectos en el sistema estudiado.
\end{itemize}

A fin de cumplir los objetivos propuestos, la tesis se divide en tres capítulos, conclusiones, recomendaciones y dos apéndices. En el Capítulo 1 se tratan todos los aspectos preliminares como las unidades y magnitudes físicas utilizadas en la tesis, se discuten las propiedades fundamentales de la condensación de Bose-Einstein y se explica el procedimiento general utilizado en los cálculos termodinámicos a través del caso a $B=0$. El Capítulo 2 se dedica al estudio del gas magnetizado de bosones vectoriales neutros en el l\'imite no relativista, mientras que en el Cap\'itulo 3 se presenta la descripción totalmente relativista de este gas. Los principales resultados y aportes de la tesis se resumen en las conclusiones, mientras que en las recomendaciones se delinean los caminos por los cuales pensamos dar continuidad a la investigación. En el apéndice A se explican los detalles del c\'alculo del calor espec\'ifico en el límite no relativista. El apéndice B recoge algunas cuestiones importantes del cálculo de la susceptibilidad magnética en el caso relativista.

\chapter{Preliminares}
\label{cap1}

Este capítulo tiene carácter introductorio. En él se presenta el sistema de unidades utilizado en la tesis, se discuten las principales características de la condensación de Bose-Einstein, y se explica la metodología a seguir para el estudio termodinámico del gas magnetizado de bosones vectoriales neutros a través de un ejemplo: un gas de bosones vectoriales neutros a campo magn\'etico cero.

\section{Unidades y magnitudes f\'isicas utilizadas }

En astrofísica y en f\'isica de part\'iculas es habitual y conveniente usar el sistema de unidades naturales (UN). Este sistema se define haciendo en todas las ecuaciones $\hbar=c=k_B=1$, donde $\hbar$ es la constante de Planck, c es la velocidad de la luz en el vac\'io y $k_B$ es la constante de Boltzman. En la tesis, todas las ecuaciones están escritas en este sistema de unidades y las magnitudes que se encuentran en ellas están expresadas en potencias de MeV:
\begin{eqnarray}\nonumber
[longitud]=[tiempo]=[masa]^{-1}=[energ\'ia]^{-1}=[temperatura]^{-1}.	
\end{eqnarray}

Los factores de conversi\'on para obtener las magnitudes en el Sistema Internacional de unidades a partir de las UN son:
\begin{eqnarray}\nonumber
1m&=&5.07\times 10^{13}MeV^{-1},\\ \nonumber
1kg&=& 5.61\times 10^{29}MeV,\\ \nonumber
1s&=& 1.52\times 10^{21}MeV^{-1},\\ \nonumber
1K&=& 8.61\times 10^{-11}MeV,\\ \nonumber
1J&=& 6.24\times 10^{12}MeV,\\ \nonumber
1T&=&10^4G=0.699\times 10^{-9}MeV^2.	
\end{eqnarray}

En los gráficos, las magnitudes se reportan o bien adimensionalizadas, o en las unidades m\'as usadas para ellas, a fin de que nuestros resultados puedan compararse fácilmente con los reportados en la literatura especializada. Por ello, la masa se expresa en MeV y la densidad de partículas en $cm^{-3}$; el campo magn\'etico se da en Gauss(G), la presi\'on en MeV/fm$^3$ y la temperatura en Kelvin (K). La densidad de masa nuclear $\rho_N$ para los neutrones corresponde a una densidad de part\'iculas $N$ de aproximadamente $10^{38}cm^{-3}$.

Todos los c\'alculos num\'ericos de la tesis se realizaron para un gas de bosones vectoriales neutros con masa $2m_n$ y momento magn\'etico $2k_n$, donde $m_n = 939,565 MeV$ y $k_n=8.65\times 10^{-5}$MeV$^{-1}$ son la masa y el momento magn\'etico del neutr\'on \cite{Zyla:2020zbs}, a fin de facilitar su comparación con lo obtenido en \cite{tesisgretel}. Es importante aclarar que aunque el apareamiento de neutrones solo tiene lugar para temperaturas $T\lesssim 10$ keV \cite{gezerlis2014pairing}, nuestro estudio no pierde fuerza por esto, ya que en todos los casos la presentaci\'on y discusi\'on de los resultados ha sido hecha de manera general y sus conclusiones son v\'alidas más allá de la part\'icula escogida.

\section{Espectro energético del gas magnetizado de bosones vectoriales neutros}

Los bosones neutros con spin uno en presencia de un campo magnético pueden ser descritos a través de una extensión del lagrangiano de Proca que incluya las interacciones entre las partículas y el campo electromagnético \cite{PhysRev.131.2326,PhysRevD.89.121701}:
\begin{eqnarray}\label{Lagrangian}
  L = -\frac{1}{4}F_{\mu\nu}F^{\mu\nu}-\frac{1}{2} \rho^{\mu\nu}\rho_{\mu\nu}
       + m^2 \rho^{\mu}\rho_{\mu}
      +i m \kappa(\rho^{\mu} \rho_{\nu}-\rho^{\nu}\rho_{\mu}) F_{\mu\nu}.
\end{eqnarray}
En la Ec.~(\ref{Lagrangian}) los índices $\mu$ y $\nu$ van de $1$ a $4$, $F^{\mu\nu}$ es el tensor del campo electromagnético y  $\rho_{\mu\nu}$, $\rho_{\mu}$ son variables independientes del campo que cumplen \cite{PhysRev.131.2326}:
\begin{equation}\label{fieldeqns}
  \partial_{\mu} \rho_{\mu\nu}-m^2 \rho_{\nu}+ 2i \kappa m \rho_{\mu} F_{\mu\nu}=0,\quad\quad
  \rho_{\mu\nu} = \partial_{\mu} \rho_{\nu} - \partial_{\nu} \rho_{\mu}.
\end{equation}
De la variación del lagrangiano con respecto al campo $\rho_{\mu}$ se obtienen las ecuaciones de movimiento, que en el espacio de los momentos pueden escribirse como:
\begin{equation}
\left((p_{\mu}^2  + m^2)\delta_{\mu\nu} -p_{\mu} p_{\nu}  - 2  i \kappa m F_{\mu \nu}\right)\rho_{\mu} = 0.
 \end{equation}
%
En lo que sigue de la tesis el campo magnético se considerará uniforme, constante y en la dirección $z$: $\textbf{B}=(0,0,B)$. Con ello, uno puede partir de la Ec.~(\ref{fieldeqns}) y obtener el hamiltoniano generalizado de Sakata-Taketani para la función de onda de seis componentes que describe el sistema de bosones vectoriales magnetizados \cite{PhysRev.131.2326, PhysRevD.89.121701} siguiendo el procedimiento descrito en \cite{PhysRev.131.2326}. El hamiltoniano es \cite{tesisgretel}:
\begin{equation}\label{hamiltonian}
\hat{H} = \sigma_3 m + (\sigma_3 + i \sigma_2) \frac{\textbf{p}^2}{2 m} -
    i \sigma_2 \frac{(\textbf{p}\cdot\textbf{S})^2}{m}
    -(\sigma_3 - i \sigma_2) \kappa \textbf{S} \cdot \textbf{B},
\end{equation}
\noindent con $\textbf{p}=(p_{\perp},p_3)$, donde $p_3$ es la componente del momentum paralela al campo magnético y $p_{\perp}=p_1^2 + p_{2}^2$ es la componente perpendicular. $\sigma_{i}$ son las matrices de Pauli\footnote{
$\begin{array}{ccc}\sigma_1=
 \left(
\begin{array}{cc}0 & 1  \\1 & 0 \end{array}
\right),
& i\sigma_2=
\left(
\begin{array}{cc} 0 & 1  \\ \text{-}1 & 0\end{array}
\right),
&\sigma_3=
\left(
\begin{array}{cc}1&0\\0&\text{-}1\end{array}
\right)
\end{array}$ \\[1pt]}
de $2\times2$, $S_{i}$ son las matrices de $3\times3$ de spin uno en una representación en la que $S_3$ es diagonal y $\textbf{S} = \{S_1,S_2,S_3\}$\footnote{
$\begin{array}{ccc} S_1=\frac{1}{\sqrt{2}}
\left( \begin{array}{ccc}
0 & 1& 0\\
1 & 0 & 1\\
0 & 1 & 0
\end{array}\right),
& S_2=\frac{i}{\sqrt{2}}
\left( \begin{array}{ccc}
0 & \text{-}1& 0\\
1 & 0 & \text{-}1\\
0 & 1 & 0
\end{array} \right),
& S_3=
\left( \begin{array}{ccc}
1 & 0& 0\\
0 & 0 & 0\\
0 & 0 & \text{-}1\end{array}\right)\end{array}$}.

A partir del hamiltoniano pueden obtenerse las ecuaciones de movimiento para el momento $\textbf{p}$ y la posición $\textbf{r}$:
\begin{eqnarray}
\frac{\partial \textbf{p}}{\partial t} = i [\hat{H},\textbf{p}], \hspace{1.5cm}\frac{\partial \textbf{r}}{\partial t} = i [\hat{H},\textbf{r}]
\end{eqnarray}
\noindent donde $[a,b] = ab-ba$ es el conmutador de $a$ y $b$. Las ecuaciones de movimiento quedan:
\begin{equation} \label{motion}
 \frac{\partial \textbf{p}}{\partial t}=\vec{0},
 \end{equation}
\begin{equation} \label{motionr}
m  \frac{\partial \textbf{r}}{\partial t}= (\sigma_3 - i \sigma_2) \textbf{p} + i \sigma_2 [\textbf{S}, \textbf{p}, \textbf{S}].
 \end{equation}
De la Ec.~(\ref{motion}) se sigue que los bosones vectoriales neutros se mueven libremente en la dirección paralela al campo así como en la perpendicular \cite{tesisgretel}. Esto es una diferencia con respecto al caso de los bosones vectoriales cargados, en el cual la componente del momento perpendicular al campo está cuantizada \cite{Elizabeth}.

Los autovalores del hamiltoniano dan lugar al espectro de los bosones:
%
\begin{eqnarray}\label{eq0.9}
\epsilon(p_3,p_{\perp},B,s)= \sqrt{m^2+p_3^2+p_{\perp}^2-2kBs\sqrt{m^2+p_{\perp}^2}},
\end{eqnarray}
\noindent donde $s=0,\pm 1$ son los autovalores del spin \cite{tesisgretel}. La componente perpendicular del momentum $p_{\perp}$ se acopla al campo magn\'etico reflejando la simetría axial que impone  al sistema el campo magn\'etico.
%
El estado b\'asico para un bos\'on neutro de spin uno se obtiene al hacer $p_3=p_{\perp}=0$ y $s=1$:
\begin{eqnarray}\label{eq0.10}
\epsilon(0,B)=\sqrt{m^2-2kBm}=m\sqrt{1-b},
\end{eqnarray}
donde $b=B/B_c$ y $B_c=m/2k$. De la expresión (\ref{eq0.10}) podemos ver que a medida que B aumenta, la energía del estado básico disminuye hasta hacerse cero en $B=B_c$, y compleja más allá de este valor. Por tanto, para $B\geq B_c$ el sistema se vuelve inestable.

La Ec.(\ref{eq0.10}) nos permite definir el momento magn\'etico efectivo de cada part\'icula como \cite{tesisgretel}:
\begin{equation}
d=-\frac{\partial \varepsilon(0,B)}{\partial B}=\frac{k m}{\sqrt{m^2-2mk B}}=\frac{k}{\sqrt{1-b}}\label{magmoment},
\end{equation}
de donde se sigue que el sistema tiene un comportamiento param\'agnetico, pues $d>0$, algo que ser\'a importante en la discusi\'on de las propiedades magn\'eticas, al igual que la divergencia de $d$ cuando $b \rightarrow 1$ $(B \rightarrow B_c)$ .

Cuando $p_3, p_{\perp}, kB<<m$ los efectos relativistas dejan de ser apreciables. En este caso, teniendo en cuenta que $\sqrt{m^2+p_{\perp}^2}\sim m+\frac{p_{\perp}^2}{2m}$ y $p_3^2+p_{\perp}^2=p^2$, la ecuación (\ref{eq0.9}) se transforma en:
\begin{eqnarray}\label{eq0.13}
\epsilon(p_3,p_{\perp},B,s)\approx\sqrt{m^2+p^2-2kBsm-2s\frac{kB\; p_{\perp}^2}{2m}}.
\end{eqnarray}
\noindent El t\'ermino $\frac{kB\; p_{\perp}^2}{2m}$ de la expresión anterior se puede despreciar ya que es de tercer orden. Luego, haciendo el desarrollo en serie de Taylor de $m\sqrt{1+(\frac{p}{m})^2-\frac{2kBs}{m}}$ hasta el segundo orden, obtenemos el siguiente espectro para el gas de bosones vectoriales neutros en el límite no relativista (NR):
\begin{eqnarray}\label{eq0.14}
\epsilon_{NR}(p,B,s)=m+\frac{p^2}{2m}-skB.
\end{eqnarray}
Para simplificar las expresiones termodinámicas que obtendremos a partir del espectro NR, vamos a reescalar la Ec.(\ref{eq0.14}) con respecto al término constante de la masa a través de la sustitución  $\epsilon\rightarrow \epsilon-m$. Esto es equivalente a hacer  $\mu\rightarrow\mu-m$ en el potencial termodinámico. La única magnitud que se ve afectada por este cambio es la densidad de energía total $E$. Pero ella se puede corregir fácilmente sumándole el t\'ermino $mN$. Finalmente, el espectro no relativista queda:
\begin{eqnarray}\label{eq0.15}
\epsilon_{NR}(p,B,s)=\frac{p^2}{2m}-skB.
\end{eqnarray}

\section{Condensado de Bose-Einstein}\label{CBE}

Una de las caracter\'isticas m\'as sobresalientes de los gases bos\'onicos es la ocurrencia de la condensaci\'on de Bose-Einstein (CBE). Este  estado de la materia fue predicho por Albert Einstein en 1924, a partir de estudios previos de Satyendra Nath Bose. Einstein  obtuvo, de manera te\'orica, que al enfriar un sistema  bosónico por debajo de cierta temperatura cr\'itica se produce una concentraci\'on de las part\'iculas en el nivel de mínima energía, al contrario de lo que ocurre para un sistema fermiónico, en el que un comportamiento similar está prohibido por el Principio de exclusión de Pauli. La ocupaci\'on del estado fundamental de un sistema cu\'antico por un número macrosc\'opico de part\'iculas es lo que se conoce como CBE. A pesar de  los numerosos intentos llevados a cabo durante más de siete décadas para producir este fen\'omeno en el laboratorio, no fue  hasta  junio de 1995 que se logró el primer condensado \cite{anderson1995observation}. 
Este se obtuvo al enfriar un vapor diluido de aproximadamente dos mil átomos de rubidio-87  por debajo de $170$ nK a través de una combinación de enfriamiento por láser y enfriamiento por evaporación magnética. En los últimos 20 a\~{n}os se han obtenido numerosos CBE en diferentes sistemas atómicos \cite{nikuni2000bose,fried1998bose,bradley1997bose,sebastian2005characteristic}. Incluso, recientemente se ha producido el condensado de Bose-Einstein en la estación espacial internacional, confirmando que los efectos  gravitatorios pueden despreciarse  en su obtenci\'on \cite{becker2018space}.

%
%

En este epígrafe resumiremos los principales supuestos teóricos que dan lugar al CBE en aras de facilitar la discusiones físicas  de las propiedades magnéticas del gas de bosones. Para el análisis partiremos de la teoría cuántica de campo a temperatura finita, ya que ella incluye todos efectos que nos interesan: la descripción relativista para el gas de bosones y la discusión de las antipartículas.

En teoría cuántica de campos a temperatura finita se denomina “carga” $\widehat{Q}$ a cualquier número cuántico conservado. La carga conservada asociada con el n\'umero de bosones se define como \cite{haber1981thermodynamics}:
\begin{equation}\label{eq1.10.1}
Q=\sum_{p}\frac{1}{e^{(\epsilon(p)-\mu)/T}-1}-\sum_{p}\frac{1}{e^{(\epsilon(p)+\mu)/T}-1}.
\end{equation}
%
El primer y segundo término de la expresión anterior corresponden al número de partículas $N^{+}$ y antipartículas $N^{-}$ respectivamente. Como podemos apreciar, $N^{+}$ y $N^{-}$ no se conservan por separado; lo que se conserva es su diferencia  $N^{+}-N^{-}$ determinada por la temperatura $T$ y el potencial qu\'imico $\mu$ del sistema \cite{haber1981thermodynamics}.

Para que $N^{+}$ y $N^{-}$ sean definidas positivas, de la expresión (\ref{eq1.10.1}) se llega a la importante conclusión de que $|\mu|\leq m$. Por otra parte, nótese que el signo de $Q$ depende de si $\mu$ toma valores positivos o negativos, y esto indica si las partículas superan en número a las antipartículas o viceversa. Algo interesante a destacar es el hecho de que no hay ninguna restricción matemática para que el número de antipartículas sea mayor que el de partículas. Sin embargo, hasta ahora las evidencias experimentales y observacionales sugieren que vivimos en un universo donde la materia prima sobre la antimateria, una preferencia que la física aún no logra explicar. El experimento ALPHA en el CERN está dedicado precisamente al estudio de las propiedades de la antimateria a fin de mejorar nuestro entendimiento de la misma \cite{bertsche2005alpha,ahmadi2017observation}.

%
Si en la Ec.(\ref{eq1.10.1}) hacemos el paso de la suma a una integral sobre $p$, la densidad de carga  o densidad de part\'iculas $N=Q/V$ se convierte en:
\begin{equation}\label{eq1.10.3}
N=\frac{1}{2\pi^2}\int_{0}^{\infty} p^2dp \bigg[\frac{1}{e^{(\epsilon(p)-\mu)/T}-1}-\frac{1}{e^{(\epsilon(p)+\mu)/T}-1}\bigg].
\end{equation}
La ecuación anterior constituye una fórmula implícita para $\mu$ como función de $N$ y $T$. Para $|\mu|<m$, la dependencia $\mu(N,T)$ siempre puede ser determinada. En cambio, si $\mu=\pm m$ el integrando de la Ec.(\ref{eq1.10.3}) diverge en  $\epsilon(p=0)=m$. En este caso hay dos posibilidades que dependen de la dimensión y el espectro del sistema \cite{beckmann1979bose}. La primera es que la integral de la Ec.(\ref{eq1.10.3}) diverja para $\mu=\pm m$; esto significa que no existe ningún par de valores de $N$ y $T$ para los cuales $\mu$ sea igual a la energía mínima del sistema. La segunda posibilidad es que la integral converja aún cuando $\mu=\pm m$, y esto significa que existen ciertos valores (c\'iticos) de $N$ y $T$ a partir de los cuales $\mu=\pm m$. En tal caso es posible demostrar que la Ec.(\ref{eq1.10.3}) solo describe a las part\'iculas en los estados excitados \cite{haber1981thermodynamics,haber1982finite}. Por tanto, la diferencia entre el valor de la Ec.(\ref{eq1.10.3}) y el número total de partículas en el sistema corresponde al n\'umero de bosones en el estado fundamental, es decir, al número de bosones condensados. 



\begin{figure}[h!]
	\centering
	\includegraphics[width=0.49\linewidth]{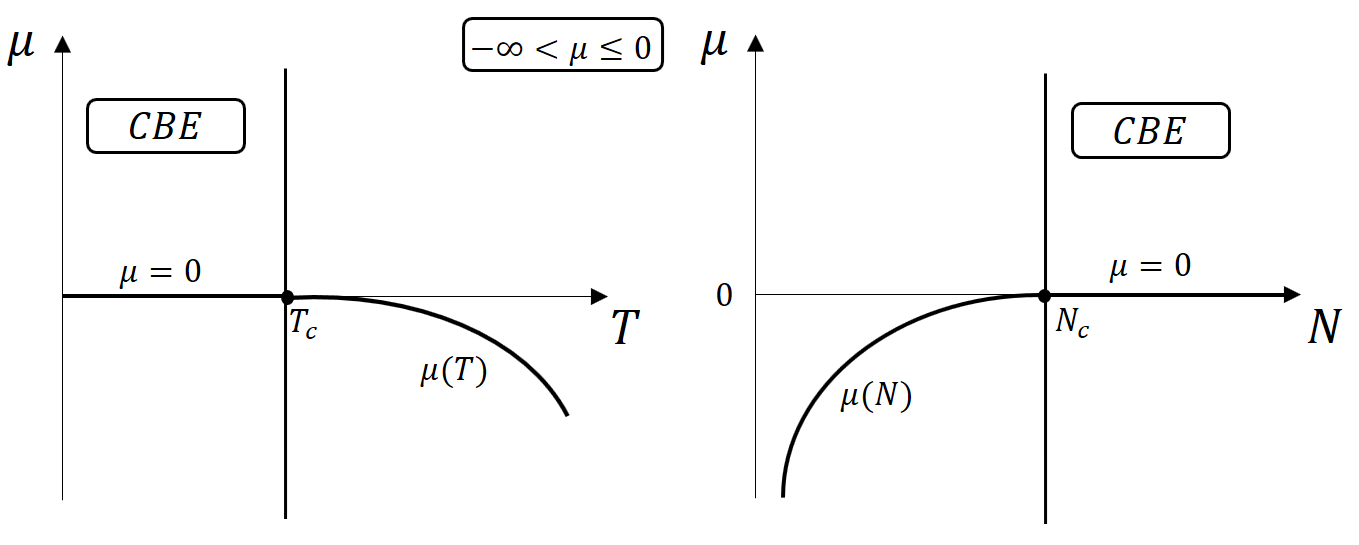}
	\includegraphics[width=0.49\linewidth]{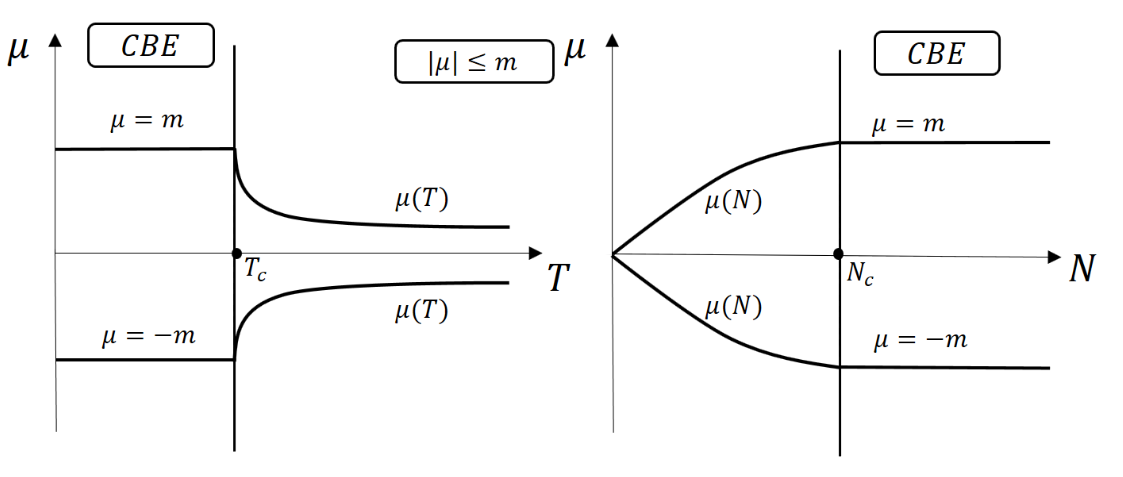}
	\caption{\label{potencialquimico1} Representaci\'on esquemática de la dependencia del potencial qu\'imico con la temperatura y la densidad de part\'iculas para un gas ideal de bosones no relativistas (panel derecho) y relativista (panel izquierdo). En todos los casos, la condición para la ocurrencia del condensado es $\mu = \pm \epsilon(p=0)$.}
\end{figure}

En los casos en que la CBE es posible, la densidad de carga puede escribirse de manera general como:
\begin{eqnarray}\label{condL}
N =
\left\{ \begin{array}{ccc}
N_{gs} + \frac{1}{2\pi^2}\int_{0}^{\infty} p^2dp \bigg[\frac{1}{e^{(\epsilon(p)-\mu)/T}-1}-\frac{1}{e^{(\epsilon(p)+\mu)/T}-1}\bigg], &  \mu=\pm m,\,\, T< T_c,\,\,N> N_c \\
& \\
\frac{1}{2\pi^2}\int_{0}^{\infty} p^2dp \bigg[\frac{1}{e^{(\epsilon(p)-\mu)/T}-1} -\frac{1}{e^{(\epsilon(p)+\mu)/T}-1}\bigg],& |\mu|<m,\,\,  T\geq T_c,\,\, N\leq N_c
\end{array} \right.,
\end{eqnarray}
donde $N_{gs}$ es el número de partículas en el estado básico. A pesar de que el CBE es comúnmente caracterizado como un fenómeno de baja temperatura, en realidad está determinado por la relación entre la temperatura y la densidad de partículas. La región condensada está delimitada por $T<T_c$ y $N>N_c $, como se sigue de la Ec.(\ref{condL}) y se ha representado en la Figura \ref{potencialquimico1}.
Si la densidad es lo suficientemente alta, la condensación puede ocurrir incluso a temperaturas relativistas $T_c>>m$, un escenario en el que las antipartículas juegan un papel muy importante, como veremos a continuación.

En el límite ultrarelativista  se cumple que $T,p>>m$ y $T>>\mu$, y lo usual es aproximar el espectro de energía a $\epsilon(p) \cong p$ y despreciar los t\'erminos que contienen a $\mu/T$ en la definición de $N$. Pero n\'otese que si se desprecia $\mu/T$ en la Ec.(\ref{eq1.10.3}) el término de las partículas se cancela con el de las antipartículas. Para no perder la información relacionada con el potencial químico es necesario realizar un desarrollo en serie de Taylor de la función $1/(e^{(p\pm \mu)/T}-1)$ alrededor de $\mu/T = 0$ hasta segundo orden. De esta forma se obtiene \cite{haber1981thermodynamics,burakovsky1996new}:
\begin{equation}\label{eq1.10.4}
|N|\cong \frac{\mu}{\pi^2 T}\int_{0}^{\infty} p^2dp \bigg[\frac{e^{p/T}}{(e^{p/T}-1)^2} \bigg] \cong \frac{\mu T^2}{3}.
\end{equation}
Por el contrario, si desde un incio no se consideran las antipartículas, al tomar el límite ultrarrelativista $N$ queda:
\begin{equation}\label{eq1.10.4b}
N\cong \frac{1}{2\pi^2}\int_{0}^{\infty} p^2dp \bigg[\frac{1}{(e^{p/T}-1)} \bigg] \cong \frac{\zeta[3]T^3}{\pi^2},
\end{equation}
\noindent donde $\zeta$ es la funci\'on zeta de Riemann. La temperatura cr\'itica $T_c$ y la densidad de part\'iculas en el estado fundamental serían:

\begin{eqnarray}\label{eq1.10.6}
T_c=
\begin{cases}
\left(\frac{3|N|}{m}\right)^{1/2},& \text{con antipart\'iculas},\\ \\
\left(\frac{\pi^2N}{\zeta[3]}\right)^{1/3},& \text{sin antipart\'iculas},
\end{cases}
\end{eqnarray}
y
\begin{eqnarray}\label{eq1.10.5}
N_{gs}=
\begin{cases}
|N|\left[1-\left(\frac{T}{T_c}\right)^2\right],& \text{con antipart\'iculas},\\ \\
N\left[1-\left(\frac{T}{T_c}\right)^3\right],& \text{sin antipart\'iculas}.
\end{cases}
\end{eqnarray}

Despreciar las antipartículas al estudiar el CBE relativista en el régimen de altas temperaturas $T>>m$ conduce a resultados incorrectos. Esto se aprecia claramente a partir de la Ec.(\ref{eq1.10.5}), pues la dependencia entre el número de partículas en el condensado y la temperatura es cuadrática cuando se tienen en cuenta las antipartículas, y cúbica cuando estas se desprecian. En el l\'imite no relativista (tratado en el pr\'oximo ep\'igrafe) $N_{gs}$ depende de $T^{3/2}$ (Ec.(\ref{eq1.15})). 

Por otra parte, la Ec.(\ref{eq1.10.6}) muestra que tener en cuenta a las antipartículas implica una dependencia entre $T_c$  y la masa del bosón. Este resultado tiene implicaciones interesantes para $m\rightarrow 0$, pues en este caso $T_c \rightarrow \infty$ y $N_0=N$. Luego, la carga neta de un gas ideal de bosones no masivo 
reside en el estado fundamental\footnote{Esto no sucede así para un gas de fotones porque en este caso la ausencia de un número cuántico conservado impide la condensación \cite{haber1981thermodynamics}.}. Como se mencionó en el epígrafe anterior, uno de los efectos del campo magnético sobre un sistema de bosones vectoriales neutros consiste en la disminución de la energía de su estado básico (Ec.(\ref{eq0.10})). En consecuencia, en los próximos capítulos veremos que al aumentar el campo magnético, aumenta la temperatura de condensación.

De la Ec.(\ref{eq1.10.6}) con antipart\'iculas también se sigue que para obtener la condensación a temperaturas relativistas $T_c>>m$, tiene que cumplirse que $N>>m^3$. Por el contrario, en el régimen no relativista lo que se cumple es $N<<m^3$, resultado que puede encontrarse en cualquier libro de física estadística estándar \cite{huang1987statistical}. En otras palabras, las altas densidades favorecen la condensación en el sentido de que esta se puede obtener a temperaturas altas.


\section{Propiedades termodinámicas del gas de bosones vectoriales neutros a $B=0$}
\label{cap2}


\subsection{Gas no relativista}

Un gas queda descrito teóricamente si se conoce su ecuaci\'on de estado, y esta siempre se puede obtener a partir del potencial termodinámico por unidad de volumen $\Omega(\mu,T)$.  Este potencial para el gas no relativista de bosones se calcula como \cite{CarlosR}:
\begin{equation} \label{eq0.1w}
\Omega(\mu,T)=- T\int_{\epsilon_{min}}^\infty d\epsilon \; g(\epsilon)\; \ln\bigg[\bigg(1-e^{\frac{\mu-\epsilon}{T}}\bigg)^{-1}\bigg],
\end{equation}
\noindent donde $g(\epsilon)$ es la densidad de estados:
\begin{equation}\label{eq1.10}
g(\epsilon)=\sum_{-s}^s \sum_{p} \delta\left[\epsilon-\frac{p^2}{2m}\right]=\frac{(2m)^{3/2}}{4 \pi^2} \epsilon^{1/2},\quad\; 0\leq\epsilon<\infty.
\end{equation}
\noindent Sustituyendo la Ec.(\ref{eq1.10}) en la Ec.(\ref{eq0.1w}) el potencial termodinámico queda:
\begin{equation}\label{eq1.2}
\Omega_{NR}(\mu,T)=-\bigg(\frac{m}{2\pi}\bigg)^{3/2}T^{5/2}g_{5/2}(z),
\end{equation}
\noindent siendo $g_{(5/2)}=\sum_{l=1}^{\infty}\frac{z^l}{l^{5/2}}$ la funci\'on polylogar\'itmica de orden $5/2$ y $z=e^{\mu/T}$ la fugacidad.

\noindent La densidad de partículas se determina derivando la Ec.(\ref{eq1.2}) con respecto a $\mu$:
\begin{equation}\label{eq1.12}
N(\mu,T,0)=N_{gs}-\left(\frac{\partial\Omega_{NR}(\mu,T)}{\partial\mu}\right)_{T}=N_{gs}+\bigg(\frac{mT}{2 \pi}\bigg)^{3/2}g_{(3/2)}(z).
\end{equation}

\begin{figure}[h!]
	\centering
	\includegraphics[width=0.55\linewidth]{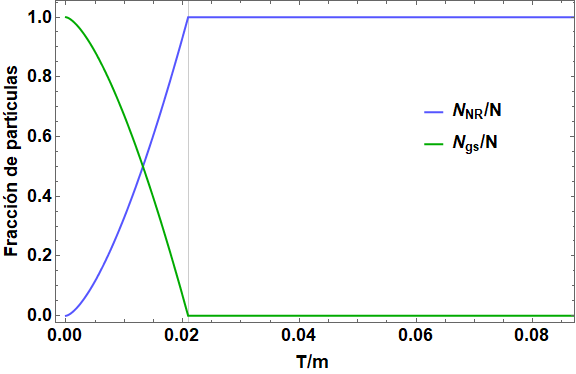}
	\caption{\label{particlefraction1} Fracción de partículas en función de la temperatura para $N=1.30 \times 10^{39}cm^{-3}$. La línea verde es la fracción de partículas en el estado fundamental. La línea azul corresponde a la fracción de partículas no condensadas.}
\end{figure}

%
\noindent Para una densidad fija, la fracci\'on de part\'iculas en el estado fundamental puede calcularse como:
\begin{eqnarray}\label{eq1.15}
\frac{N_{gs}}{N}=
\begin{cases}
0 & T>T_c\\
1-\left(\frac{T}{T_c}\right)^{3/2} & T<T_c.
\end{cases}
\end{eqnarray}
\noindent Los parámetros críticos $T_c(N)$ y $N_c(T)$, Ec.(\ref{eq1.15b}), se obtienen evaluando la Ec.(\ref{eq1.12}) en la condición del condensado, que para el gas ideal de bosones no relativistas es $\mu =0$:
\begin{eqnarray}\label{eq1.15b}
T_c(N)=\frac{2 \pi}{m}\bigg(\frac{N}{g_{(3/2)}(1)}\bigg)^{2/3},\;\;\;\;N_c(T)=\bigg(\frac{mT}{2 \pi}\bigg)^{3/2}g_{(3/2)}(1).
\end{eqnarray}
La Figura \ref{particlefraction1} muestra la fracción de partículas en los estados excitados $N_{NR}/N$ y en el estado fundamental $N_{gs}/N$ para  $N=1.30 \times 10^{39}cm^{-3}$.
En la misma se observa que el número de partículas en el condensado es cero siempre que la temperatura est\'e por encima de $T_c$, mientras que por  debajo de $ T_c $ las part\'iculas comienzan a caer en el estado de m\'inima energ\'ia.  A $T = 0$, se obtiene lo que se denomina el condensado puro, es decir, todas las partículas están en el estado fundamental.

\begin{figure}[h!]
	\centering
	\includegraphics[width=0.55\linewidth]{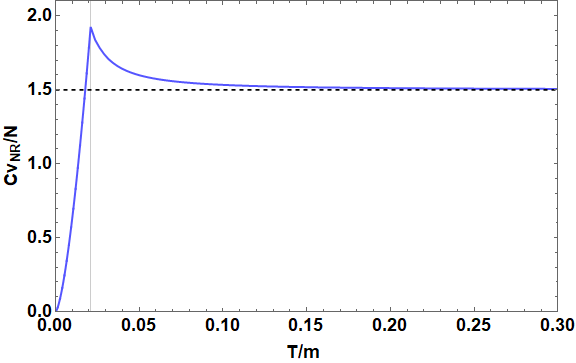}	
	\caption{\label{Cv1} Calor espec\'ifico por part\'icula como funci\'on de la temperatura para el gas de bosones no relativistas a campo magnético cero.}
\end{figure}

El resto de las magnitudes termodinámicas, dígase la presi\'on, la energ\'ia interna, la entrop\'ia y la la capaciad calor\'ifica  por unidad de volumen tienen la forma:
\begin{eqnarray}\label{energia2}
P_{NR}(\mu,T)&=&-\Omega_{NR}(\mu,T),\\\label{entropia}
S_{NR}(\mu,T)&=&-\left(\frac{\partial\Omega_{NR}(\mu,T)}{\partial T}\right)_{\mu}=-\frac{5}{2}\frac{\Omega_{NR}}{T}-N\ln z,\\\label{energia}
E_{NR}(\mu,T)&=&\Omega_{NR}(\mu,T)+TS_{NR}(\mu,T)-\mu \left(\frac{\partial\Omega_{NR}(\mu,T)}{\partial\mu}\right)_{T}=-\frac{3}{2}\Omega_{NR}(\mu,T),
\end{eqnarray}
\begin{eqnarray}\label{Cv}
C_{V_{NR}}(\mu,T)=\left(\frac{\partial E_{NR}(\mu,T)}{\partial T}\right)_{\mu}=
\begin{cases}
\frac{15}{4} \left(\frac{m}{2\pi}\right)^{3/2}T^{5/2} g_{(5/2)}(z),&\text{CBE},\\
\frac{15}{4}N\frac{g_{(5/2)}(z)}{g_{(3/2)}(z)}-\frac{9}{4}N\frac{g_{(3/2)}(z)}{g_{(1/2)}(z)},& \text{Gas Libre}.
\end{cases}
\end{eqnarray}
Como la energ\'ia del estado fundamental para el gas no relativista es cero $\epsilon(p=0)=0$, las part\'iculas que se encuentran en este estado no contribuyen a la energ\'ia interna del sistema, ni a la presi\'on, ya que no tienen energ\'ia cin\'etica ($p=0$). De la Ec.(\ref{entropia}) se obtiene que $S_{NR} = 0$ para $T = 0$, lo que est\'a en concordancia con con la Tercera Ley de la Termodinámica \cite{huang1987statistical}. Esto significa que la fase condensada tiene entropía cero.
La Figura \ref{Cv1} muestra el calor específico por partícula en función de la temperatura. Del gráfico podemos ver el pico bien definido que indica la transición de fase al estado condensado. Por debajo de la temperatura cr\'itica el calor espec\'ifico decrece como $T^{3/2}$ y a altas  temperaturas tiende a $3/2$.

Para que el gas quede completamente descrito termodinámicamente se hace necesario conocer la dependencia del potencial qu\'imico $\mu(N,T)$ con respecto a la temperatura y la densidad de part\'iculas  fuera del estado condensado. Hallar $\mu(N,T)$ es equivalente a hallar $z(N,T)$. El potencial qu\'imico puede tomar valores entre $-\infty<\mu\leq0$, luego la fugacidad est\'a acotada entre $0<z\leq1$, y la condensaci\'on ocurre para $z=1$. La expresión de la fugacidad en la fase condensada y de gas libre es:
\begin{equation}\label{eq1.13}
z(N,T,0)=
\begin{cases}
1, & \text{CBE},\\
N-\bigg(\frac{mT}{2 \pi}\bigg)^{3/2}g_{(3/2)}(z)=0, & \text{Gas Libre}.
\end{cases}
\end{equation}

\subsection{Gas relativista}
\label{MTBR}
El espectro del gas de bosones relativistas no magnetizados se obtiene haciendo $B=0$ en la expresi\'on (\ref{eq0.9}). En ausencia de campo magn\'etico la simetría esférica es recuperada, por tanto $p_3^2+p_{\perp}^2=p^2$ y el espectro queda: $\sqrt{p^2+m^2}$. El potencial termodinámico y la densidad de estados son en este caso:
\begin{equation} \label{eq0.1x}
\Omega^{\pm}(\mu,T,0)=-T\int_{\epsilon_{min}}^\infty d\epsilon \; g_{R}(\epsilon)\; \ln\left[\left(\left(1-e^{\frac{\mu-\epsilon}{T}}\right)\left(1-e^{\frac{-\mu+\epsilon}{T}}\right)\right)^{-1}\right],
\end{equation}
\begin{eqnarray}\label{eq7}
g_{R}(\epsilon)=\sum_{s}\sum_{\vec{p}}\delta \bigl[\epsilon-\sqrt{p^2+m^2}\bigr]
=\frac{4\pi V}{(2\pi)^3}(2s+1)\;2\epsilon\;\sqrt{\epsilon^2-m^2},\;\;\; \epsilon\geq m.
\end{eqnarray}
\noindent Haciendo uso de:
\begin{equation}
ln(1-x)=-\sum_{n=1}^{\infty}\frac{x^n}{n} \, ,
\end{equation}
y de:
\begin{equation}
\int_u^{\infty}x(x^2-u^2)^{\nu-1}e^{-\alpha x}dx=2^{\nu-\frac{1}{2}}(\sqrt{\pi})^{-1}u^{\nu+\frac{1}{2}}\Gamma(\nu)K_{\nu+\frac{1}{2}} \, ,
\end{equation}
el potencial termodinámico queda \cite{su2008thermodynamic}:
\begin{equation}\label{eq1.22}
\Omega^{\pm}(\mu,T)=-\bigg(\frac{m^2T^2}{2\pi^2}\bigg)\sum_{n=1}^{\infty}\frac{z^n+z^{-n}}{n^2}K_2(nm/T),
\end{equation}
\noindent donde $K_{\alpha}$ es la funci\'on de MacDonald de orden $\alpha$. En la Ec.(\ref{eq1.22}) se han tomado en cuenta las antipartículas a través del término $z^{-n}$, mientras que $z^{n}$ corresponde a las partículas.

La dependencia del potencial qu\'imico para el caso relativista es diferente a la no relativista, como se muestra esquem\'aticamente en el panel derecho de la Figura \ref{potencialquimico1}. Aquí los valores posibles de $\mu$ están acotados entre $0<\mu\leq m$, luego la fugacidad podrá tomar valores entre $1< z\leq e^{m/T}$, y el estado condensado se alcanza cuando $z=e^{m/T}$.

Las densidades de partículas ($N^{+}$) y antipartículas ($N^{-}$) que se obtienen  luego de derivar el potencial termodinámico  Ec.(\ref{eq1.22}) con respecto a $\mu$ son:
\begin{eqnarray}\label{eq1.23}
N^{+}(\mu,T)=\frac{m^2T}{2 \pi^2}\sum_{n=1}^{\infty}\frac{z^{n}}{n}K_2(nm/T),
\end{eqnarray}
\begin{eqnarray}\label{eq1.24}
N^{-}(\mu,T)=\frac{m^2T}{2 \pi^2}\sum_{n=1}^{\infty}\frac{z^{-n}}{n}K_2(nm/T).
\end{eqnarray}
\noindent Donde ahora $N_{ext}=N^{+}(\mu,T)-N^{+}(\mu,T)$ es la densidad carga en los estados excitados. Las ecuaciones  (\ref{eq1.23}) y (\ref{eq1.24}) est\'an en concordancia con las obtenidas en\cite{elmfors1995condensation,beckmann1979bose}. Como se muestra en el panel derecho de la Figura \ref{potencialquimico1}, cuando  $T\rightarrow \infty$ el potencial químico tiende a  cero $\mu\rightarrow 0$, lo que implica $N^{-}\rightarrow N^{+} y \;\;N_{ext}\rightarrow0$, esto es lo que se conoce como vac\'io caliente \cite{perez2006negative}.

\begin{figure}[h!]
	\centering
	\includegraphics[width=0.49\linewidth]{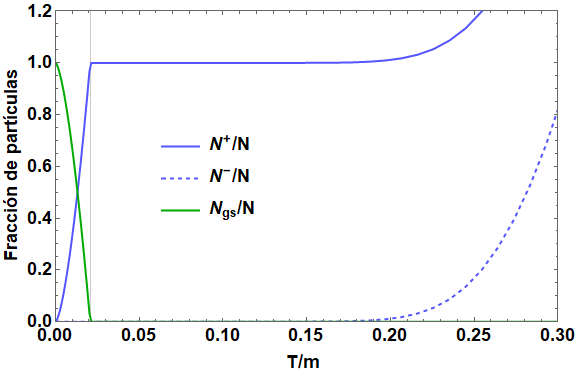}
	\includegraphics[width=0.49\linewidth]{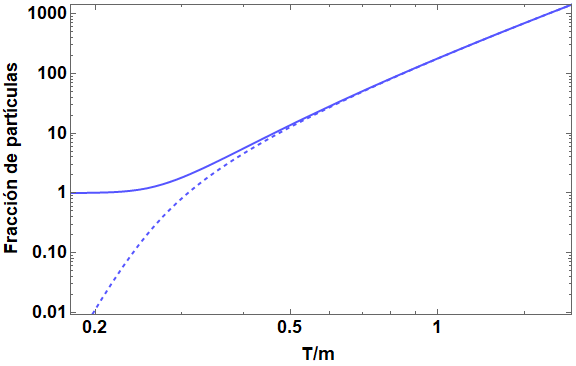}
	\caption{\label{particlefraction} Fracción de partículas, antipartículas y partículas en el condensado en función de la temperatura. La línea verde es la fracción de partículas en el estado fundamental. La línea azul continua corresponde a la fracción de partículas y la línea azul discontinua corresponde a la fracción de antipartículas.}
\end{figure}

La Figura \ref{particlefraction} muestra la fracci\'on de part\'iculas  $N^+(\mu,T)/N$, antipart\'iculas $N^-(\mu,T)/N$ y partículas en el estado fundamental $N_{gs}/N$ para $N=1.30 \times 10^{39}cm^{-3}$.

El panel derecho de la Figura \ref{particlefraction} es la continuaci\'on en temperatura del panel izquierdo; esta separación se realizó para que los efectos que queremos resaltar puedan observarse mejor. Como podemos ver en el panel izquierdo, $N_{gs}=0,\; \forall T>T_c$, lo que corresponde al gas libre, mientras que $N_{gs} \neq 0,\; \;\forall T<T_c$, lo que corresponde a un estado donde el gas libre y el condensado coexisten. Por otro lado, la contribución de las antipartículas comienza a ser importante para $ T \simeq \frac{1}{5}m $, mientras que para $ T\gtrsim m$ las fracciones de partículas y antipartículas ya son del mismo orden. Este es el vac\'io caliente que se mencion\'o antes. Por tanto, en este rango de temperaturas la presencia de las antipartículas podr\'ia tener una influencia importante en algunas propiedades termodinámicas del gas, tales como la presión y la magnetización.

\begin{figure}[h!]
	\centering
	\includegraphics[width=0.55\linewidth]{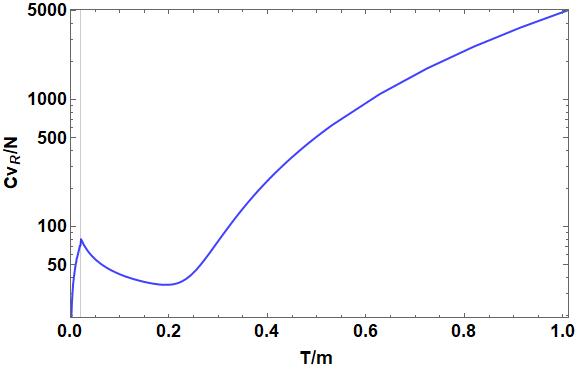}	
	\caption{\label{Cv2} Calor espec\'ifico por part\'icula como funci\'on de la temperatura para el gas de bosones  relativistas.}
\end{figure}

\noindent Las expresiones (\ref{energia2})-(\ref{Cv}) para el gas relativista se transforman en:
\begin{eqnarray}\label{eq1.25}
P(\mu,T)&=&-\Omega^{\pm}(\mu,T),\\
S(\mu,T)&=&-\frac{2}{T}\Omega^{\pm}(\mu,T)-\frac{\mu}{T}(N^{+}-N^{-})\\
\nonumber
&+&\frac{m^3}{4\pi^2}\sum_{n=1}^{\infty}\frac{z^n+z^{-n}}{n}[K_1(nm/T)+K_3(nm/T)],\\ \label{eq1.26}
E(\mu,T)&=&-\Omega^{\pm}(\mu,T)\\\nonumber
&+&\frac{m^3T}{4\pi^2}\sum_{n=1}^{\infty}\frac{z^n+z^{-n}}{n}[K_1(nm/T)+K_3(nm/T)],\\
C_V(\mu,T)&=&S(\mu,T)+\bigg\{\frac{m^3}{4\pi^2}\sum_{n=1}^{\infty}\frac{z^n+z^{-n}}{n}[K_1(nm/T)+K_3(nm/T)]\\ \nonumber
&+&\frac{m^4}{8\pi^2T}\sum_{n=1}^{\infty}(z^n+z^{-n})[K_0(nm/T)+2K_2(nm/T)+K_4(nm/T)]\\ \nonumber
&-&\frac{\mu}{T}\frac{m^3}{4\pi2}\sum_{n=1}^{\infty}(z^n-z^{-n})[K_1(nm/T)+K_3(nm/T)]\bigg\}.
\end{eqnarray}
La Figura \ref{Cv2} muestra el calor específico por partícula en función de la temperatura para el gas relativista.
Del gráfico podemos ver el pico bien definido que indica la transición de fase al estado condensado al igual que en el caso no relativista. La principal diferencia con respecto al caso no relativista es que a altas temperaturas $C_{V}$ ya no tiende a $3/2$, sino que aumenta  debido a la presencia de las antipartículas.

En lo que resta de la tesis, los c\'alculos aqu\'i presentados a $B=0$ se extender\'an a $B\neq0$ para el caso no relativista (Cap\'itulo 2) y el caso relativista (Cap\'itulo 3). Las expresiones a $B=0$ ser\'an usadas para comprobar la validez de las expresiones con campo magn\'etico haciendo $B\rightarrow0$ en ellas. Adem\'as, el caso $B=0$ ser\'a incluido en los gr\'aficos para estudiar el efecto del campo magn\'etico en las magnitudes termodin\'amicas.

    \chapter{Estudio termodin\'amico del gas magnetizado de bosones vectoriales neutros en el l\'imite no relativista}
\label{cap2}

En este capítulo se estudian las propiedades termodinámicas de un gas magnetizado de bosones vectoriales neutros en el l\'imite no relativista. Entre ellas, la condensación de Bose-Einstein, que depende de todos los parámetros involucrados en el problema: temperatura, densidad de partículas y campo magnético; las ecuaciones de estado, y las propiedades magnéticas del gas. Este estudio recoge los resultados originales de la autora publicados en \cite{suarez2019non}.

\section{Potencial termodinámico}

Partiendo del espectro (\ref{eq0.15}), $\epsilon(p)=\vec{p}^{\:2}/2m -sk B$, la densidad de estados del GBVN no relativista en presencia de un campo magn\'etico externo constante y uniforme $\vec{B}=(0,0,B)$ se determina por:
\begin{eqnarray}\label{eq2.1}\nonumber
g_{NR}(\epsilon)&=&
\sum_{s=-1,0,1} \sum_{\vec{p}} \delta\left(\epsilon-\frac{\vec{p}^{\:2}}{2m}+sk B\right)\\
&=& \sum_{\vec{p}} \delta\left(\epsilon-\frac{\vec{p}^{\:2}}{2m}-k B\right)+
\delta\left(\epsilon-\frac{\vec{p}^{\:2}}{2m}\right)+
\delta\left(\epsilon-\frac{\vec{p}^{\:2}}{2m}+k B\right).
\end{eqnarray}
\noindent N\'otese que en (\ref{eq2.1}) tenemos tres posibles estados de spin: $ s=-1,0,1 $, ya que los bosones son vectoriales. Usando el cambio de  sumatoria a integral $\sum_{ \vec{p}}=\frac{V}{(2\pi )^3}\int_{-\infty}^\infty dp_{x} \int_{-\infty}^\infty dp_{y}  \int_{-\infty}^\infty dp_{z}, $ y pasando a coordenadas esféricas se obtiene:
\begin{eqnarray}\label{eq2.3.0}
g_{NR}(\epsilon)= \frac{mV}{2\pi^2} \left[\int_0^{\infty} p^2 \delta(\epsilon-\frac{\vec{p}^{\:2}}{2m}+kB)dp+\int_0^{\infty} p^2 \delta(\epsilon-\frac{\vec{p}^{\:2}}{2m})dp+\int_0^{\infty} p^2 \delta(\epsilon-\frac{\vec{p}^{\:2}}{2m}-kB)dp \right].
\end{eqnarray}
Con el cambio de variables $x_i=p^2/2m+ikB$, donde $i=0,\pm1$ y la propiedad $\int_0^{\infty} f(x)\delta(x-a)dx=f(a)$ finalmente la densidad de estados queda:
\begin{eqnarray}\label{eq2.3}
g_{NR}(\epsilon)= \frac{ mV}{2\pi^2} \left[ \sqrt{2m(\epsilon_{-}-k B)}+ \sqrt{2m \epsilon_{0}} + \sqrt{2m(\epsilon_{+}+k B)} \right],
\end{eqnarray}
\begin{eqnarray}\nonumber
\;\;\;kB\leq\epsilon_{-}<\infty,\;\;\;\;0\leq\epsilon_{0}<\infty,\;\;\;\;-kB\leq\epsilon_{+}<\infty.
\end{eqnarray}
\noindent Con la densidad de estados obtenida en la Ec.(\ref{eq2.3}) se calcula el potencial termodinámico \cite{CarlosR}:
\begin{equation} \label{eq0.1}
\Omega_{NR}(\mu,T,B)=-T\int_{\epsilon_{min}}^\infty d\epsilon \; g_{NR}(\epsilon)\; \ln\left[\left(1-e^{\frac{\mu-\epsilon}{T}}\right)^{-1}\right],
\end{equation}
\begin{eqnarray}\nonumber
\Omega_{NR}(\mu,T,B)= \frac{ mV}{2\pi^2}\bigg(\int_{kB}^{\infty}d\epsilon\sqrt{2m(\epsilon-k B)}\ln[1-ze^{-\epsilon/T}]+
\int_{0}^{\infty}d\epsilon\sqrt{2m\epsilon}\ln[1-ze^{-\epsilon/T}] \bigg)\\\label{eqpotencial}
+\int_{-kB}^{\infty}d\epsilon\sqrt{2m(\epsilon+k B)}\ln[1-ze^{-\epsilon/T}].
\end{eqnarray}
N\'otese que en esta expresi\'on los l\'imites de integraci\'on son diferentes de acuerdo con el valor del spin. Luego de integrar por la energ\'ia, obtenemos el potencial termodin\'amico NR dividido en tres sumandos, uno por cada estado de spin:
\begin{eqnarray}\label{eq2.4a}
\Omega_{NR}(\mu,T,B)&=&\Omega_{-_{NR}}(\mu,T,B)+\Omega_{0_{NR}}(\mu,T,B)+\Omega_{+_{NR}}(\mu,T,B),\\\label{eq2.4b}
\Omega_{NR}(\mu,T,B)&=&-\bigg(\frac{m}{2 \pi}\bigg)^{3/2}T^{5/2}[g_{5/2}(z_{-})+g_{5/2}(z)+g_{5/2}(z_{+})]
\end{eqnarray}

\noindent donde $z_{i}=z e^{i\frac{\kappa B}{T}}$ con $i=\pm1$.

\section{Condensaci\'on de Bose-Einstein y propiedades magn\'eticas} 

Comenzaremos el estudio de la condensación de Bose-Einstein calculando la densidad de partículas
\begin{equation}\label{eq2.5}
N = N_{gs}(T,B) +N_{NR}(\mu,T,B),
\end{equation}
\noindent donde $N_{gs}$ representa las partículas en el condensado, y $N_{NR} = -d\Omega_{NR}/d\mu$ las part\'iculas en los estados excitados. Despu\'es de hacer la derivada con respecto a $\mu$ en (\ref{eq2.4a}), la densidad de part\'iculas toma la forma:

\begin{eqnarray}\label{eq2.6}
N &=&N_{gs}(T,B) + N_{-}(\mu,T,B)+N_{0}(\mu,T,B)+N_{+}(\mu,T,B),\\
N &=& N_{gs}(T,B) +\bigg(\frac{mT}{2 \pi}\bigg)^{3/2}[g_{3/2}(z_{-})+ g_{3/2}(z)+ g_{3/2}(z_{+})] .\label{eq2.7}
\end{eqnarray}

Las curvas de los parámetros críticos $N_c(T, b)$ y $T_c(N, b)$ se obtienen evaluando la Ec.(\ref{eq2.7}) en la condición del condensado. Para el gas no relativista esta condición es $\mu =-kB$, por tanto:

\begin{eqnarray}\label{eq2.6b}
	N_c=\bigg(\frac{mT_c}{2 \pi}\bigg)^{3/2}[g_{3/2}(e^{-2kB/T_c})+ g_{3/2}(e^{-kB/T_c})+ g_{3/2}(1)].
\end{eqnarray}

\begin{figure}[h!]
	\centering
	\includegraphics[width=0.49\linewidth]{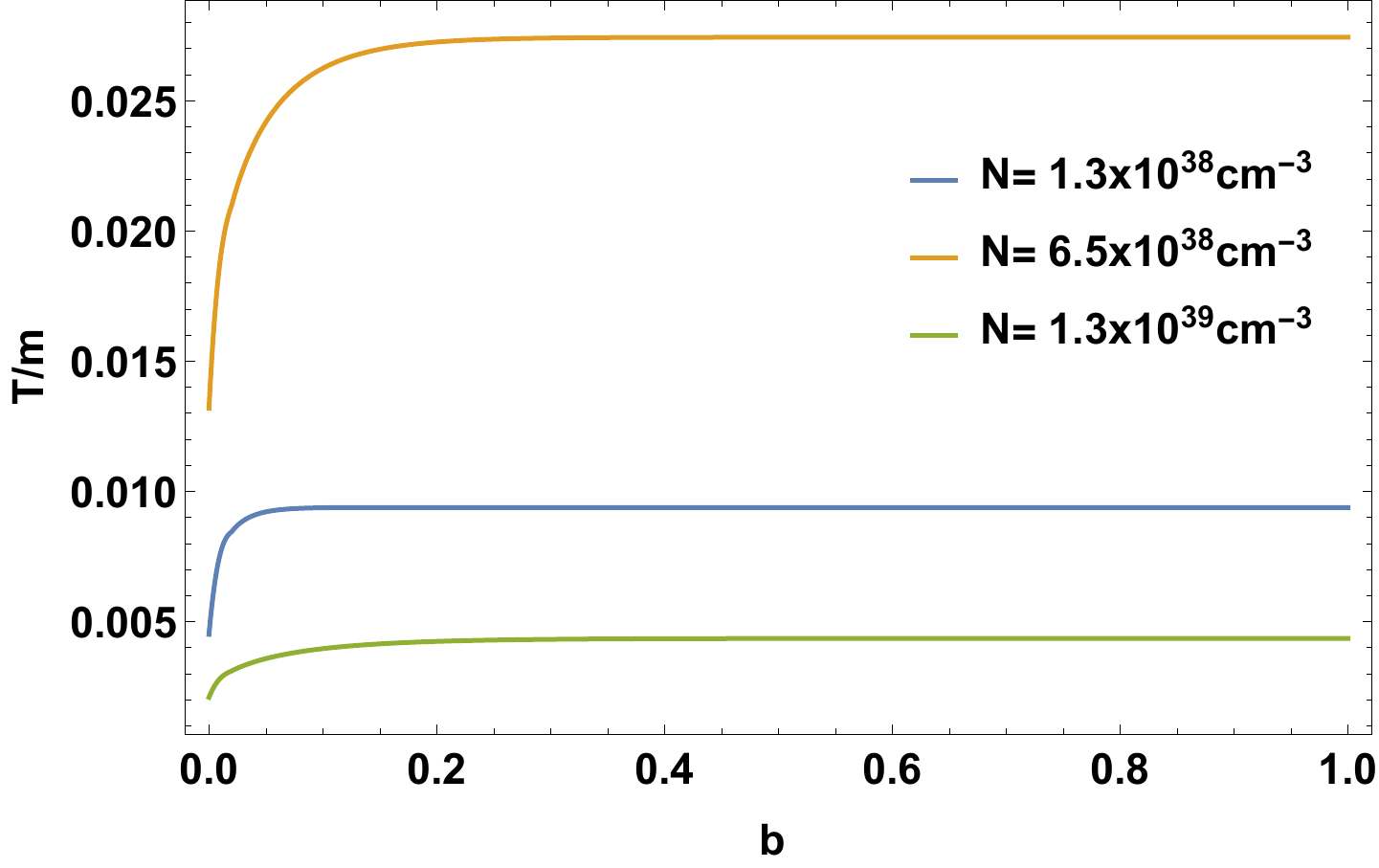}
	\includegraphics[width=0.49\linewidth]{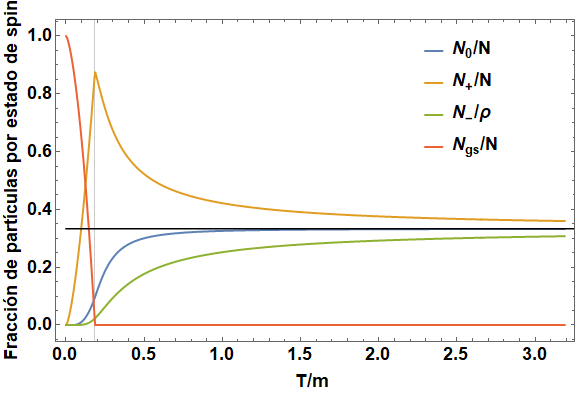}
	\caption{\label{f2.1}Panel izquierdo: temperatura crítica  en función del campo magnético para varios valores de la densidad de partículas. Panel derecho: densidad de partículas por estado de spin en función de la temperatura para $ N=1.30\times10^{39}$cm$^{-3}$ y $b=0.5$.}
\end{figure}

A partir de la Ec.~(\ref{eq2.6b}), la temperatura cr\'itica del condensado fue calculada numéricamente en función de $ N $ y $ B $ como se muestra en el panel izquierdo en la Figura \ref{f2.1}. 

El panel izquierdo de la Figura \Ref{f2.1} muestra cómo crece la temperatura crítica a partir de su valor en $ B = 0 $:
\begin{equation}\label{eq2.8a}
T_{c}(0)= \frac{2\pi}{m} \left( \frac{N}{3 g_{3/2}(1)}  \right)^{2/3},
\end{equation}
\noindent hasta la saturación cuando $B \rightarrow \infty$:
\begin{eqnarray}\label{eq2.8b}
T_{c}(\infty)=\frac{2\pi}{m} \left( \frac{N}{ g_{3/2}(1)}  \right)^{2/3}.
\end{eqnarray}
Estos valores extremos de la temperatura crítica del CBE están en concordancia con los obtenidos en \cite{yamada1982thermal}. 

El comportamiento asintótico de $ T_c $ con el aumento del campo magnético nos brinda  información valiosa sobre la influencia de $ B $ en el CBE de las partículas no relativistas. En la región de campo magnético débil ($B\rightarrow 0$)  el incremento de $ B $ aumenta $ T_c $ de manera notable, lo que lleva al sistema a la condensación. Pero cuando el campo magnético es fuerte, sus cambios apenas afectan la temperatura crítica. Por tanto, a densidad de partículas fija, el efecto del campo magnético en el CBE es aumentar la temperatura crítica, de manera que cuanto más débil es el campo, más sensible es el sistema al mismo. N\'otese también que aunque la temperatura crítica del CBE aumenta con la densidad de partículas, la raz\'on entre los dos valores extremos $ T_c (\infty) /T_c(0) = \sqrt[3]{9} $ es constante.

\begin{figure}[h!]
	\centering
	\includegraphics[width=0.55\linewidth]{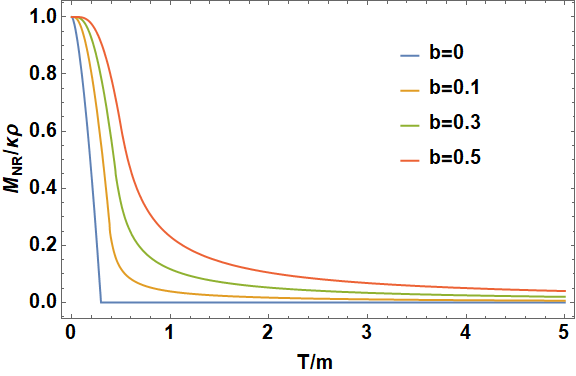}
	\caption{\label{f2.2} Magnetizaci\'on como funci\'on de la temperatura para $N=1.30\times10^{39}$cm$^{-3}$ y diferentes valores del campo magn\'etico.}
\end{figure}

Es interesante observar el comportamiento de la fracción de partículas en el estado fundamental $N_{gs}/N $ y por estado de spin $ N_{i} / N $, $ {i=-,0,+} $ en función de la temperatura (panel derecho de la Figura~\Ref{f2.1}). En la región de alta temperatura, $ T \gg m $, $ N_{gs}=0 $ y $ N_{i} / N\rightarrow 1/3 $ para todo $ i $, debido a que esta regi\'on es dominada por el desorden térmico. Cuando $ T $ disminuye, el efecto del campo magnético comienza a notarse y la fracción de partículas con spin alineado al campo magn\'etico, (es decir $ N_{+} / N$), se convierte en la dominante. Este comportamiento continúa a través de la región de baja temperatura $ T \ll m $ siendo el próximo cambio apreciable cuando $ T=T_c $. En este punto, la fracción de partículas en el estado fundamental deja de ser cero y aumenta con la  disminución de la temperatura hasta que alcanza su valor m\'aximo, $ 1 $, en $ T = 0 $ (donde $ N_{i} / N=0 $ para todo $ i $). Dado que en el estado fundamental  $s=1$, se espera que para un gas de bosones vectoriales que ha estado magnetizado, exista una magnetización distinta de cero incluso cuando $B \rightarrow 0$. Para verificar esto, calculemos la magnetización del gas:
\begin{equation}\label{eq2.9}
M_{NR}(\mu,T,B)=k N_{gs} -\left(\frac{\partial \Omega_{NR}(\mu,T,B)}{\partial B}\right) =  k(N_{gs} + N_{+}- N_{-}),
\end{equation}
\noindent y grafiqu\'emosla en la Figura~\ref{f2.2} a $N$ fija y varios valores de $B$.

Las curvas en este gráfico concuerdan con las del panel derecho de la Figura~\ref{f2.1}: $ M \rightarrow 0 $ para altas $ T $, mientras que $ M \rightarrow k N $ cuando $ T \rightarrow 0 $. La característica m\'as sobresaliente de la Figura~\ref{f2.2} es que esto sucede también para $ B = 0 $, es decir, el gas muestra una magnetización espontánea que en este caso no se debe a una interacción entre el spin de las partículas sino al condensado, pues para $ B = 0 $, $ M \neq 0 $ solo cuando $ T <T_c $. Este fenómeno, conocido como ferromagnetismo de Bose-Einstein \cite{yamada1982thermal,simkin1999magnetic}, es un resultado de gran relevancia astrofísica, pues la magnetización espontánea podría funcionar como fuente de campo magnético en el interior de los objetos compactos.

La aparición de una magnetización espontánea también se puede obtener mediante la sustitución directa de $ B = 0 $ en la ecuación~(\ref{eq2.9}), que da $ M (B = 0) = k N_ {gs} $, y su conexión con el condensado puede estudiarse a partir del comportamiento del calor espec\'ifico y la susceptibilidad magn\'etica del gas. 

Para $T<T_c$ el calor espec\'ifico por unidad de volumen tiene la forma: 
\begin{eqnarray} \label{eq2.10a}
C_{V_{NR}}(\mu=-kB,T,B)=\frac{3}{2}S_{NR}-\frac{3}{2}\frac{kB}{T}[N_{+}-N_{-}]+2\left(\frac{kB}{T}\right)^2\left(\frac{mT}{2\pi}\right)^{3/2}g_{1/2}((e^{-2kB/T})),
\end{eqnarray}
\noindent mientras que para $T>T_c$:
\begin{eqnarray}\label{eq2.10b}
C_{V_{NR}}(\mu,T,B)=\frac{15}{4}N\frac{g_{5/2}(z_{-})+g_{5/2}(z)+g_{5/2}(z_{+})}{g_{3/2}(z_{-})+g_{3/2}(z)+g_{3/2}(z_{+})}-\frac{9}{4}N
\frac{\frac{[g_{3/2}(z_{-})]^{2}}{g_{1/2}(z_{-})}+\frac{[g_{3/2}(z)]^{2}}{g_{1/2}(z)}+\frac{[g_{3/2}(z_{+})]^{2}}
	{g_{1/2}(z_{+})}}{g_{3/2}(z_{+})+g_{3/2}(z)+g_{3/2}(z_{+})}.
\end{eqnarray}
\noindent En ambos casos el procedimiento para obtener $C_{V_{NR}}$ resulta bastante largo; sus detalles se muestran en el Ap\'endice A.

Por otra parte, para calcular la susceptibilidad magn\'etica basta derivar (\ref {eq2.9}) con respecto a $B$. Teniendo en cuenta que $ \frac {\partial}{\partial B}g_{\nu} (z_{i}) = \frac{k} {T}g_{\nu-1}(z_{i}) $, se obtiene:
\begin{equation}\label{eq2.10c}
\chi_{NR}(\mu,T,B)=
\begin{cases}
2 \frac{k^{2}}{T\lambda^{3}}g_{1/2}(e^{-2kB/T}), & \text{CBE},\\
\frac{k^{2}}{T\lambda^{3}}[g_{1/2}(z_{+})+g_{1/2}(z_{-})],& \text{Gas Libre.}
\end{cases}
\end{equation}

\begin{figure}[h!]
	\centering
	\includegraphics[width=0.49\linewidth]{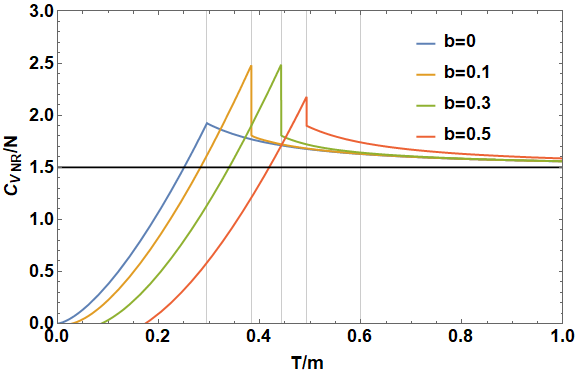}
	\includegraphics[width=0.49\linewidth]{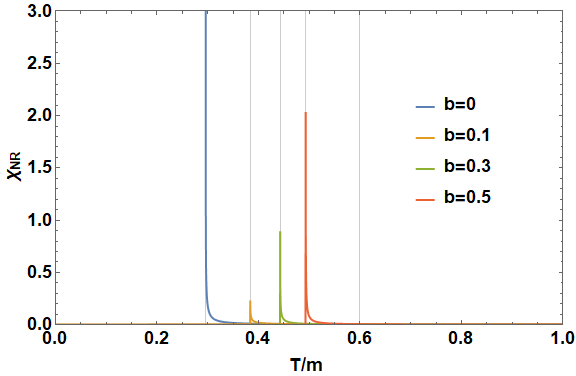}
	\caption{\label{susceptibilities} Calor específico y susceptibilidad magn\'etica  en función de la temperatura para varios valores del campo magnético y $N=1$.$30\times10^{39}$cm$^{-3}$.}
\end{figure}

La Figura \ref{susceptibilities} muestra el calor específico y la susceptibilidad magn\'etica en función de la temperatura para varios valores del campo magnético. 
Los máximos de ambas magnitudes ocurren a $T_c$, señalando así la transición de fase al condensado ferromagnético. A medida que $B$ aumenta, $T_c$ también y ello se refleja en un movimiento en la posición de los máximos de $C_{V_{NR}}$ y $\chi_{NR}$ de $T_{0}$ hacia $T_{\infty}$. En el l\'imite de alta temperatura $T>>m$ el calor específico tiende a  $3/2$ como predice la teoría clásica de los gases.

Una cuestión interesante, también relacionada con el ferromagnetismo de Bose-Einsten, aparece cuando hacemos el campo magn\'etico cero en la Ec.(\ref{eq2.10c}). En la región no condensada $z_{-}=e^{-2kB/T}$, en consecuencia si $ B = 0 \Longrightarrow z_{-} = 1 $. La función polilogar\'itmica de orden $1/2$ diverge en $1$, por tanto, en toda esta regi\'on la susceptibilidad magn\'etica est\'a indefinida:
\begin{equation}\label{eq48}
\chi(\mu,T,0)=
\begin{cases}
\infty, & \text{CBE},\\
2 \frac{2\kappa^{2}}{T\lambda^{3}}g_{1/2}(z), & \text{Gas Libre.}
\end{cases}
\end{equation}

\section{Ecuaciones de estado}

Como consecuencia de  la ruptura de la simetría rotacional $SO(3)$ producida por el campo magnético, el tensor de energía-momento de un gas cuántico magnetizado se vuelve anisotrópico y la presión se separa en dos componentes, una paralela, $ P_{\parallel}$, y la otra perpendicular, $P_{\perp}$, al eje magnético (ecuaciones (\ref{presionpara}) y (\ref{presionper})) \cite{Chaichian:1999gd}. Luego las EdE para el GBVN magnetizado en el límite no relativistas son:
\begin{eqnarray}\label{EoS}
E_{NR}(\mu,T,B) &=&\Omega_{NR}(\mu,T,B) + \mu N -T S_{NR}(\mu,T,B), \label{energia}\\
P_{\parallel_{NR}}(\mu,T,B)&=&  -\Omega_{NR}(\mu,T,B), \label{presionpara} \\
P_{\perp_{NR}}(\mu,T,B)&=& -\Omega_{NR}(\mu,T,B) -{\mathcal M}_{NR}(\mu,T,B) B. \label{presionper}
\end{eqnarray}
\noindent La entrop\'ia del gas bajo estudio es:
\begin{eqnarray}\label{eq2.11}
S_{NR}(\mu,T,B)\hspace{-0.1cm}=\hspace{-0.1cm}-\left(\frac{\partial\Omega_{NR}(\mu,T,B)}{\partial T}\right)_{\mu,B}\hspace{-0.35cm}= -\frac{5}{2}\frac{\Omega_{NR}(\mu,T,B)}{T} - \mu \frac{N_{NR}(\mu,T,B)}{T} - \kappa B \frac{N_+ - N_-}{T}.
\end{eqnarray}
\noindent Combinando la Ec.~(\ref{eq2.4b}) y las Ecs.~(\ref{eq2.7}) y (\ref{eq2.11}), llegamos a la densidad de energía:
\begin{eqnarray}\label{eq2.12}
E_{NR}(\mu,T,B)=-\frac{3}{2} \Omega_{NR}(\mu,T,B)- kB[N_{+}-N_{-}],
\end{eqnarray}

\begin{figure}[h]
	\centering
	\includegraphics[width=0.49\linewidth]{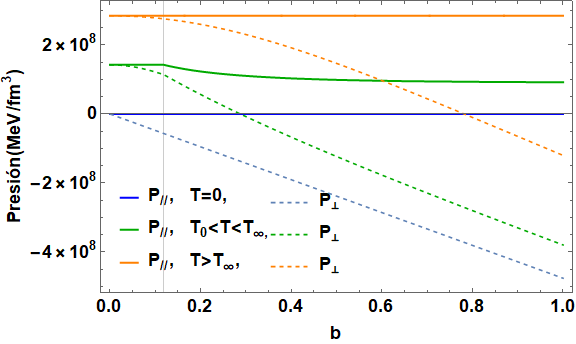}
	\includegraphics[width=0.49\linewidth]{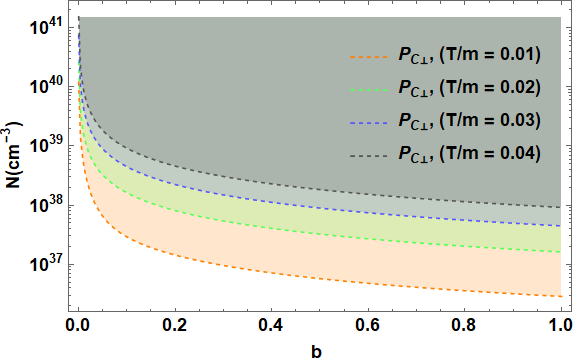}
	\caption{\label{f2.3} Panel izquierdo: presiones paralela y perpendicular en función del campo magnético para diferentes valores de temperatura y $ N=1$.$30\times 10^{39}$cm$^{-3} $. Panel derecho: diagrama de fases para el colapso magnético transversal para varios valores de la temperatura; las líneas discontinuas representan la solución de $ P_{\perp} (T, B, N) = 0 $ a temperatura fija; para cada temperatura, la región de inestabilidad ha sido sombreada.}
\end{figure}

Las EdE, Ecs.~(\Ref{presionpara}), (\ref{presionper}) y (\ref{eq2.12}), pueden usarse para modelar objetos compactos magnetizados cuya composición incluya bosones magnetizados que admitan una descripción no relativista. Sin embargo, debe tenerse en cuenta que en dependencia de la temperatura, el campo magnético y la densidad de partículas, la presión perpendicular puede ser negativa.
El panel izquierdo de la Figura~\Ref{f2.3} muestra las presiones en función del campo magnético para $N=1.3\times 10^{39}$cm$^{-3} $ y varios valores de la temperatura. Para campo magnético cero, $ P_{\parallel} = P_{\perp}$ y el sistema es isotrópico. Si $B\ne 0 $ la diferencia entre las presiones aumenta cuando disminuye la temperatura o aumenta el campo magnético. A partir de este gráfico es evidente que el campo magnético apenas afecta la presión paralela, mientras que su contribución es muy importante para la perpendicular. Desde un punto de vista microscópico, esto se debe a que el campo magnético disminuye el momento perpendicular de las partículas, pero no afecta el paralelo. Macroscópicamente, este efecto se expresa en el término sustractivo $ -M_{NR}B $ (tenga en cuenta que $ M_{NR}> 0 $) que aparece en la presión perpendicular. Es por eso que $ P_{\perp}$ siempre es menor que $P_{\parallel} $ y disminuye hasta valores negativos con el aumento del campo magnético. Dado que el efecto de una presión perpendicular negativa es empujar las partículas hacia el eje magnético, esto puede interpretarse como que el sistema se vuelve inestable. Este tipo de inestabilidad se ha observado previamente en otros gases cuánticos magnetizados y se conoce como colapso magnético transversal \cite{Chaichian:1999gd, Aurora2003EPJC,Felipe:2002wt,Elizabeth,Quintero2017AN}.

El panel derecho de la Figura~\Ref{f2.3} muestra el diagrama de fases del colapso magnético en el plano del campo magnético y la densidad de partículas, para varios valores de la temperatura. Las líneas discontinuas representan la solución de $ P_{\perp} (T, B, N) = 0 $ a una temperatura fija. El gas es estable en la región por debajo de las líneas e inestable sobre ellas. En $T\neq0 $ y $ B=0 $, se necesita un número infinito de partículas para que el sistema se vuelva inestable, un resultado esperado ya que el campo magnético es la causa de la inestabilidad. Cuando $ T = 0 $ y $ B\neq0 $, $ P_{\parallel} = 0 $ y $P_{\perp}=-M_{NR}B $ siempre es negativa, es decir, el estado condensado puro es inestable para cualquier valor del campo magnético distinto de cero. En el caso de que ambos, la temperatura y el campo magnético, sean diferentes de cero, la disminución de la temperatura favorece el colapso, así como el aumento del campo magnético. Además, la Figura \ref{f2.3} también muestra que el colapso se ve favorecido por el aumento de la densidad de partículas, ya que cuanto más denso es el gas, mayores son $ M_{NR} $ y $ T_c $. En consecuencia, para una temperatura y un campo magnético fijos, el aumento de $N $ disminuye la presión térmica $ -\Omega_{NR} (\mu, T, B) $ y aumenta la presión magnética $ -M_{NR}B $, llevando a $ P_{\perp} $ a tomar valores negativos. En este sentido, el colapso impone un límite superior a las densidades de partículas que pueden existir dentro de estrellas compactas con un campo magnético dado. No obstante, es importante señalar que dada la composición heterogénea de las estrellas, la aparición o no del colapso magnético dependerá de las presiones y, por lo tanto, de la respuesta magnética de todas las especies presentes.

    \chapter{Estudio termodin\'amico del gas magnetizado de bosones vectoriales neutros relativistas}
\label{cap3}

Este capítulo est\'a dedicado a las propiedades termodinámicas del gas relativista de bosones vectoriales neutros en presencia de un campo magn\'etico constante y uniforme. En él se obtienen las expresiones analíticas de las magnitudes termodinámicas a toda temperatura y  se investiga la influencia de las antipart\'iculas y el campo magnético en la condensación de Bose-Einstein, en la magnetización y en las ecuaciones de estado del sistema. Al mismo tiempo, se estudian la validez del límite no relativista discutido en el capitulo anterior y del límite de baja temperatura tomado en \cite{tesisgretel}. Este cap\'itulo también recoge resultados originales de la autora, algunos de los cuales ya se encuentran publicados en \cite{de2019bose}.

\section{Extensi\'on a toda temperatura. Magnitudes termodinámicas}

El potencial termodin\'amco por unidad de volumen del gas de bosones vectoriales neutros relativistas en presencia de un campo magnético externo constante y uniforme $\vec{B}=(0,0,B)$, fue previamente obtenido en \cite{tesisgretel}. 
El potencial termodinámico del gas relativista, $\Omega^{\pm}(\mu,T,b)$, puede separarse en dos términos: 
\begin{equation}\label{eq3.1}
\Omega^{\pm}(\mu,T,b)=\Omega_{st}(\mu,T,b)+\Omega_{vac}(b) \, .
\end{equation}
En la expresi\'on anterior $\Omega_{st}(\mu,T,b)$ representa la contribucci\'on estadística, que puede ser escrita en funci\'on de la contribuci\'on de las part\'iculas y antipart\'iculas, $\Omega_{st}(\mu,T,b)=\Omega^{+}_{st}(\mu,T,b)+\Omega^{-}_{st}(\mu,T,b)$, siendo:
%

\begin{eqnarray}\label{eq3.2a}
\Omega_{st}^{+}(\mu,T,b)  =  -  \sum_{s} \sum_{n=1}^{\infty} \frac{z^{ n}}{n^2}\! \bigg\{\frac{y_0^2T^2}{2 \pi^2}K_2(ny_0/T)
+  \frac{\alpha n T}{2 \pi^2} \int_{y_0}^{\infty}\frac{x^2}{\sqrt{x^2+\alpha^2}}K_1(nx/T)\bigg\} \, , 
\end{eqnarray}
\begin{eqnarray}\label{eq3.2b}
\Omega_{st}^{-}(\mu,T,b)  =  -  \sum_{s} \sum_{n=1}^{\infty} \frac{z^{-n}}{n^2}\! \bigg\{\frac{y_0^2T^2}{2 \pi^2}K_2(ny_0/T)
+  \frac{\alpha n T}{2 \pi^2} \int_{y_0}^{\infty}\frac{x^2}{\sqrt{x^2+\alpha^2}}K_1(nx/T)\bigg\} \, , 
\end{eqnarray}
\noindent donde $y_0=m\sqrt{1-sb}$ y $\alpha=mbs/2$. El segundo t\'ermino de la Ec.~(\ref{eq3.1}) es la contribución del vacío y se escribe:
\begin{eqnarray}\label{eq3.3}
\Omega_{vac}(b) & = & -\frac{m^4}{288\pi}(b^2(66-5b^2)-3(6-2b-b^2)(1-b)^2
 \ln(1-b)\\ && -3(6+2b-b^2)(1+b)^2\ln(1+b)). \nonumber 
\end{eqnarray}
Como se puede apreciar de la Ec.(\ref{eq3.3}), la contribución de vacío se anula cuando $b=0$, es por ello que no se mencionó al realizar  el estudio del GBVN a campo cero en el epígrafe \ref{MTBR}. 

El estudio llevado a cabo en \cite{tesisgretel} toma el l\'imite $T<<m$ en la Ec.~(\ref{eq3.1}) para luego obtener las magnitudes termodinámicas. El potencial termodinámico estadístico y las ecuaciones de estado que se obtienen al hacer el l\'imite de baja temperatura son:
\begin{eqnarray}\label{eq3.4a}
\Omega_{st}^{T\ll m}(\mu',T,b)=\frac{(m\sqrt{1-b})^{3/2}T^{5/2}}{2^{1/2}\pi^{3/2}(2-b)} g_{5/2}(e^{\mu'/T}), 
\end{eqnarray}
\begin{eqnarray}\label{eq3.4b}
N^{T\ll m}(\mu',T,b)=\frac{(m\sqrt{1-b})^{3/2}T^{3/2}}{2^{1/2}\pi^{3/2}(2-b)} g_{3/2}(e^{\mu'/T}), 
\end{eqnarray}
\begin{eqnarray}\label{eq3.4c}
E^{T\ll m}(\mu',T,b)=m\sqrt{1-b}\;N+\Omega_{vac}(b)-\frac{3}{2}\Omega_{st}^{T\ll m}(\mu',T,b)+T\frac{\partial \mu'}{T},
\end{eqnarray}
\begin{equation}\label{eq3.4cc}
P^{T\ll m}_{\parallel}(\mu',T,b)=-\Omega^{T\ll m}_{st}(\mu',T,b)-\Omega_{vac}(b),
\end{equation}
\begin{equation}\label{eq3.4ccc}
P^{T\ll m}_{\perp}(\mu',T,b)=P^{T\ll m}_{\parallel}(\mu',T,b)-bB_c\;\mathcal{M}^{T\ll m}(\mu',T,b),
\end{equation}
\begin{equation}\label{eq3.13}
\mathcal{M}^{T\ll m}(\mu',T,b)=\frac{kN}{\sqrt{1-b}}-\frac{kT^{3/2}}{2^{3/2}\pi^2(1-b)^{3/2}}g_{3/2}(e^{\mu'/T})+\mathcal{M}_{vac}(b),
\end{equation} 
\noindent con:
\begin{eqnarray}\label{eq3.4d}
\mu'\simeq - \frac{\zeta(3/2)T}{4\pi}\bigg(1-\bigg(\frac{T_c^{T\ll m}}{T}\bigg)^{3/2}\bigg)\Theta(T-T_c^{T\ll m}),
\end{eqnarray}
\noindent y:
\begin{eqnarray}\label{eq3.4f}
T_c^{T\ll m}= \frac{1}{m\sqrt{1-b}}\bigg(\frac{2^{1/2}\pi^{3/2}(2-b)N}{\zeta(3/2)}\bigg)^{2/3}.
\end{eqnarray}
\noindent En la Ec.(\ref{eq3.4d}), $\Theta(x)$ es la función paso unitario de Heaviside y $\mathcal M_{vac}(b)$ viene dada por la Ec.(\ref{eq3.8}).

A diferencia de lo que se hizo en \cite{tesisgretel}, en los próximos epígrafes de la tesis partiremos de las expresiones generales Ecs.~(\ref{eq3.2a}) y (\ref{eq3.2b}) para obtener las magnitudes termodinámicas a toda temperatura.

\section{Antipart\'iculas} 

\begin{figure}[h!]
	\centering
	\includegraphics[width=0.55\linewidth]{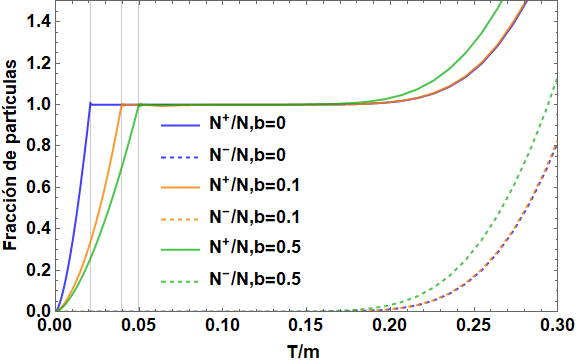}
	\caption{\label{f3.01} Fracción de partículas y antipartículas no condensadas en función de la temperatura para varios valores de campo magn\'etico.  Las líneas continuas corresponden a la fracción de partículas y las discontinuas a la de antipartículas. Las líneas horizontales se\~{n}alan la temperatura cr\'itica $T_c(b)$.  }
\end{figure}

La consecuencia más importante de usar las Ecs.~(\ref{eq3.2a}) y (\ref{eq3.2b}) en lugar de su l\'imite de baja temperatura es que en ellas se conserva la contribución de las antipartículas. Para tener una idea de cómo la presencia del campo magnético afecta la aparición de las antipartículas, estudiaremos la densidad de partículas relativista, que se escribe: $N=N_{gs}+N_{st}(\mu,T,b)$, donde $N_{gs}$ es la densidad de part\'iculas en el estado fundamental y $N_{st}(\mu,T,b)=N^{+}(\mu,T,b)-N^{-}(\mu,T,b)$ es la derivada con respecto a $\mu$ de los  potenciales termodinámicos (\ref{eq3.2a}) y (\ref{eq3.2b}), y representa la densidad total de partículas en los estados excitados:
\begin{eqnarray}\label{eq5}
N_{st}(\mu,T,b) = \sum_{s} \sum_{n=1}^{\infty}\frac{z^{ n}-z^{ -n}}{n} \bigg\{\frac{y_0^2T}{2 \pi^2} K_2(ny_0/T)
+ \frac{\alpha n}{2 \pi^2} \int_{y_0}^{\infty}\frac{x^2}{\sqrt{x^2+\alpha^2}}K_{1}(nx/T)\bigg\} \, .
\end{eqnarray}
%
%
\noindent La Figura \ref{f3.01} muestra la fracci\'on de part\'iculas  $N^+(\mu,T,b)/N$ y antipart\'iculas $N^-(\mu,T,b)/N$  no condensadas para $N=1.30 \times 10^{39}cm^{-3}$ y varios valores de campo magn\'etico.
Como se vio en el Cap\'itulo 1, la densidad de antipartículas comienza a ser apreciable para temperaturas tales que $T\gtrsim m/5$ (Figura \ref{particlefraction}). La nueva información aportada por la Figura \ref{f3.01} es que el campo magnético favorece la aparición de las antipartículas. No obstante, para que este efecto sea apreciable se requieren campos magnéticos cercanos al crítico (nótese que las curvas para $b=0$ y $b=0.1$ son prácticamente iguales). 

%
\section{Condensaci\'on de Bose-Einstein}

Para estudiar el efecto del campo magnético y de las antipart\'iculas, así como las consecuencias de tomar el l\'imite de baja temperatura, en la condensación de  Bose-Einstein, compararemos la curva cr\'itica $N_c(T,b)$  que se obtiene a partir de la expresi\'on (\ref{eq5}) con las obtenidas en el l\'imite no relativista (Cap\'itulo 2) y en el l\'imite de baja temperatura estudiado en \cite{tesisgretel}. A fin de facilitar la lectura de los gráficos, utilizaremos la siguiente nomenclatura: (R) para referirnos al gas relativista a toda temperatura, (NR) para la aproximaci\'on no relativista y (BT) para el l\'imite de baja temperatura. 

La curva $N_c(T, b)$ se obtiene evaluando la densidad de partículas en la condición del condensado. Para el gas relativista esta condición es $\mu = m \sqrt{1-b}$, por tanto:
\begin{eqnarray}\label{eq3.5}
N_c(T,b)=\hspace{-0.1cm}\sum_{s} \sum_{n=1}^{\infty}\frac{e^{nm\sqrt{1-b}/T}-e^{-nm\sqrt{1-b}/T}}{n}\bigg\{\frac{y_0^2T}{2 \pi^2} K_2(ny_0/T)
+ \hspace{-0.1cm}\frac{\alpha n}{2 \pi^2} \int_{y_0}^{\infty}\hspace{-0.3cm}\frac{x^2}{\sqrt{x^2+\alpha^2}}K_{1}(nx/T)\bigg\}.
\end{eqnarray}

\begin{figure}[h!]
	\centering
	\includegraphics[width=0.49\linewidth]{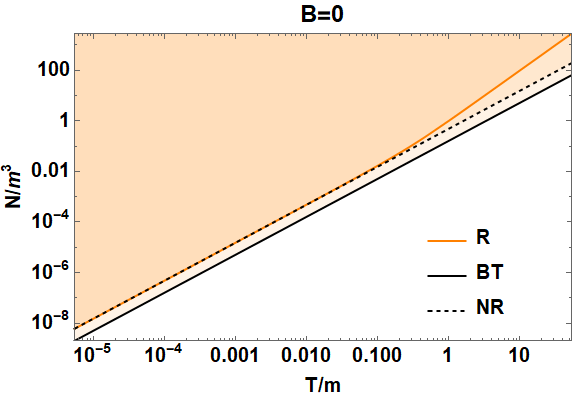}
	\includegraphics[width=0.49\linewidth]{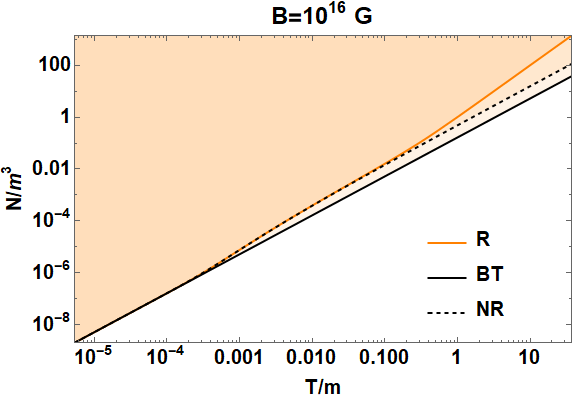}
	\caption{\label{f3.1} Diagrama de fases del condensado. La región blanca corresponde al estado de gas libre. La región sombreada corresponde al condensado. La líneas se\~{n}alan las curvas cr\'iticas $N_c(T,b)$ para diferentes descripciones del gas de bosones vectoriales neutros.}
\end{figure}

La Figura \ref{f3.1} muestra el diagrama de  fases del condensado en el plano $N-T$ para $B=0$ y $B=10^{16}$ G. 
A B=0 hay una diferencia muy notable entre el límite de baja temperatura y los otros dos casos. Esto se debe a que al tomar el l\'imite $T<<m$ en el potencial termodin\'amico, se desprecian los términos correspondientes a los estados de spin $s=0,-1$, y el único t\'ermino que sobrevive es el del estado $s=1$. Como consecuencia, una vez tomado el l\'imite de baja temperatura en presencia de campo magnético no es posible recuperar las expresiones de campo cero haciendo $b=0$ en las Ecs.(\ref{eq3.4a}-\ref{eq3.13}). Por tanto, es de esperar que este límite solo funciones bien a campos magnéticos relativamente fuertes. En este sentido, las expresiones para toda temperatura obtenidas en la tesis, Ec.(\ref{eq5}-\ref{eq8}), tienen la fortaleza de que no implican ninguna pérdida de información relacionada con el spin. Por otra parte, para $B=0$ la curva relativista y no relativista se separan alrededor de $T \sim m$, indicando la importancia de las ant\'iparticulas a esas temperaturas.
 
A $B=10^{16}$ G la curva cr\'itica en el l\'imite de baja temperatura comienza a diferenciarse de las otras dos para $T\gtrsim 10^{-3}m$, demostrando que para este valor de campo magn\'etico existe una regi\'on de temperatura en donde el l\'imite $T<<m$ s\'i es apropiado. Al igual que para campo cero, las curvas relativista y no relativista se separan alrededor de $T \sim m$. La Figura \ref{f3.1} aporta dos valores relevantes de temperatura: $T\sim10^{-3}m$, que señala el momento en que los efectos de la temperatura comienzan a ser relevantes en el gas magnetizado, y $T\sim m$ que marca el punto en que los efectos relativistas (la presencia de las antipartículas) deben tomarse en cuenta. Por \'ultimo, vale la pena se\~{n}alar que los resultados mostrados en la Figura 3.2 concuerdan con los del epígrafe \ref{CBE}, donde se mostr\'o que para que el gas de Bose se condense a temperaturas relativistas $T_c>>m$ se debe cumplir  $N_c>>m^3$. 

\begin{figure}[h!]
	\centering
	\includegraphics[width=0.55\linewidth]{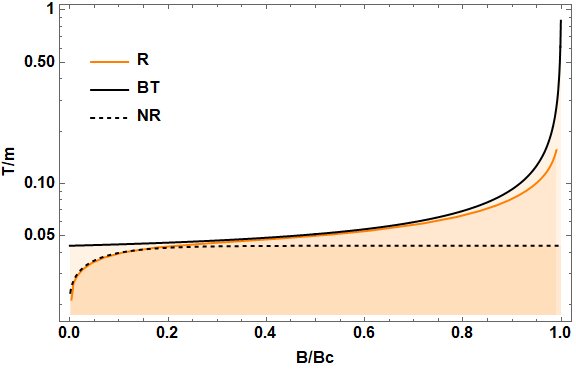}
	\caption{\label{f3.2} Temperatura cr\'itica del CBE como funci\'on del campo magnético para $N=1.30 \times 10^{39}cm^{-3}$. La región blanca corresponde al estado de gas libre. La región sombreada  corresponde al estado condensado. Las l\'ineas corresponden a las curvas cr\'iticas $T_c(b)$ para diferentes descripciones del gas de bosones vectoriales neutros.}
\end{figure}

Estudiemos ahora con mayor profundidad c\'omo afecta el campo magn\'etico la validez de los l\'imites NR y BT. Para ello fijemos la densidad de part\'iculas en $N=1.30 \times 10^{39}cm^{-3}$ y grafiquemos la temperatura cr\'itica del condensado como funci\'on del campo magnético (Figura \ref{f3.2}). 
La Figura \ref{f3.2} muestra que el comportamiento asintótico de la $T_c$ no relativista con el aumento del campo magn\'etico constituye una diferencia importante entre este y los casos relativistas. En estos últimos, la temperatura crítica siempre aumenta con $b$ y diverge cuando $b\rightarrow1$ (es decir cuando $ B\rightarrow B_c $). El comportamiento distinto de la curva no relativista a campo fuerte es una consecuencia de
suponer $kB<<m$ para la obtención del espectro no relativista, lo que significa que en este caso el campo magnético se considera débil. Por tanto, es natural que en el límite NR no se describa bien la región de campo fuerte.

Por otra parte, para $B\lesssim 0.2B_c$ la curva del l\'imite de baja temperatura comienza a discrepar de las curvas relativista y no relativista. Esto era de esperarse, pues ya sabemos que este l\'imite no funciona bien a medida que $b\rightarrow 0$. Sin embargo, cuando el campo magnético es lo suficientemente fuerte para que los efectos de la temperatura no sean los dominantes, el comportamiento del gas en el límite de baja temperatura no difiere del obtenido para toda temperatura. Como ya habíamos dicho antes, al tomar el límite de baja temperatura se desprecian los términos correspondientes a los estados de spin $s=0,-1$ y el único estado que sobrevive es el $s=1$. Esto es análogo a hacer una aproximación de campo fuerte, por lo que es lógico que el límite de baja temperatura no describa correctamente la región de campo débil. Por lo tanto, otra fortaleza de la  tesis consiste en que las expresiones para toda temperatura obtenidas en ella describen correctamente tanto la región de campo débil como la de campo fuerte. Finalmente, n\'otese que hacer $b=0$ en la Ec.(\ref{eq3.4f}) es análogo a hacer $b\rightarrow\infty$ en la expresi\'on de la $T_c$ no relativista Ec.(\ref{eq2.8b}), pues al aumentar el campo magnético se favorece al estado de spin $s=1$.

\section{Propiedades magn\'eticas}

La magnetización del gas relativista puede escribirse como la suma de tres términos: la magnetización estad\'istica $\mathcal{M}_{st}(\mu,T,b)$, la del vacío $\mathcal{M}_{vac}(b)$ y la de las partículas que se encuentran en el estado básico $\mathcal{M}_{gs}(T,b)$:
\begin{equation}\label{eq3.6}
\mathcal{M}^{\pm}(\mu,T,b)=\mathcal{M}_{st}(\mu,T,b)+\mathcal{M}_{vac}(b)+\mathcal{M}_{gs}(T,b).
\end{equation}
\noindent La contribución estadística se calcula derivando $\Omega_{st}(\mu,T,b)$ con respecto al campo magnético. Para ello reescribamos el potencial termodinámico estadístico de la siguiente manera:
\begin{equation}\label{eqb1}
\Omega_{st}(\mu,T,b)=-\sum_{s} \sum_{n=1}^{\infty} \!\frac{z^n+z^{-n}}{n}\bigg\{\frac{A_1(T,b)}{n}+A_2(T,b)\bigg\},
\end{equation}
\noindent donde:  
\begin{equation}\label{eqb2}
A_1(T,b)=\frac{m^2(1-bs)T}{2\pi^2}K_2(nm\sqrt{1-bs}/T),
\end{equation}
\begin{equation}\label{eqb3}
A_2(T,b)=\frac{mbsT}{4\pi^2}\int_{m\sqrt{1-bs}}^{\infty}\frac{x^2}{\sqrt{x^2+(mbs/2)^2}}K_2(nx/T)\;dx.
\end{equation}
Entonces, la contribución estadística a la magnetización queda:
\begin{equation}\label{eqb4}
\mathcal{M}_{st}(\mu,T,b)=-\frac{\partial \Omega_{st}(\mu,T,b)}{\partial b}\frac{\partial b}{\partial B}=\frac{2k}{m}\sum_{s} \sum_{n=1}^{\infty} \!\frac{z^n+z^{-n}}{n}\bigg\{\frac{1}{n}\frac{\partial A_1(T,b)}{\partial b}+\frac{\partial A_2(T,b)}{\partial b}\bigg\}.
\end{equation} 
\noindent La derivada de la expresión (\ref{eqb2}) es:
\begin{equation}\label{eqb5}
\frac{\partial A_1(T,b)}{\partial b}=\frac{m^2y_0\;nsT}{4\pi^2}K_1(ny_0/T),
\end{equation} 
\noindent mientras que para derivar la integral de la expresión (\ref{eqb3}) utilizaremos la siguiente propiedad:
\begin{equation}\label{eqb8}
\frac{\partial}{\partial x}\int_{a(x)}^{b(x)}f(x,t)dt=\int_{a(x)}^{b(x)}\frac{\partial}{\partial x}[f(x,t)]dt+f[b(x)]b'(x)-f[a(x)]a'(x).
\end{equation} 
\noindent Con ella la derivada de (\ref{eqb2}) queda:
\begin{eqnarray}\label{eqb9}
\frac{\partial A_1(T,b)}{\partial b}=\frac{msT}{4\pi^2}\int_{y_0}^{\infty}\frac{x^2}{\sqrt{x^2+\alpha^2}} K_1(nx/T)\;dx
-\frac{\alpha^3T}{2\pi^2b}\int_{y_0}^{\infty}\frac{x^2}{(x^2+\alpha^2)^{3/2}}K_1(nx/T)\;dx\\
+\frac{\alpha^2y_0T}{\pi^2b(2-bs)}K_1(ny_0/T),
\end{eqnarray} 
\noindent con lo cual se obtiene, despu\'es de simplificar:
\begin{equation}\label{eq3.7}
\mathcal{M}_{st}(\mu,T,b)=\hspace{-0.1cm}\sum_{s} \sum_{n=1}^{\infty} \!\frac{z^n+z^{-n}}{n}\bigg\{\frac{my_0ks\;T}{\pi^2(2-bs)}K_1(ny_0/T)+\hspace{-0.1cm}\frac{ksT}{2\pi^2}\int_{y_0}^{\infty}\hspace{-0.3cm}\frac{x^4}{(x^2+\alpha^2)^{3/2}}K_1(nx/T)\;dx\bigg\}.
\end{equation}
\noindent De esta expresi\'on es evidente que las part\'iculas/antipart\'iculas en el estado con spin perpendicular al campo magnético, es decir con $s=0$, no contribuyen a la magnetización del gas. 

La contribución del vacío se calcula derivando la Ec.(\ref{eq3.3}) con respecto al campo magnético:
\begin{equation}\label{eq3.8}
\mathcal{M}_{vac}(b)=-\frac{k m^3}{72 \pi}(7b(b^2-6)-3(2b^3-9b+7)\log(1-b)-3(2b^3-9b-7)\log(1+b)).
\end{equation}

Para obtener la contribución de las partículas que se encuentran en el estado fundamental multiplicamos el momento magnético efectivo de cada partícula $d=k/\sqrt{1-b}$, que aumenta con la intensidad del campo magnético, por la densidad de partículas que se encuentran en ese estado: 
\begin{equation}\label{eq3.9}
\mathcal{M}_{gs}(T,b)=\frac{k}{\sqrt{1-b}}N_{gs}(T,b).
\end{equation}
\noindent Sustituyendo las Ecs.(\ref{eq3.7}), (\ref{eq3.8}) y (\ref{eq3.9}) en la Ec.(\ref{eq3.6}) obtenemos la magnetizaci\'on total del gas relativista a toda temperatura.

La Figura \ref{f3.3} muestra la magnetizaci\'on como funci\'on de la temperatura para las diferentes descripciones del gas de bosones vectoriales neutros. En el gr\'afico hemos incluido a modo de referencia la contribuci\'on del vac\'io. Los c\'alculos se hicieron para   $N=1.30 \times 10^{39}cm^{-3}$ y dos valores de campo magn\'etico,  $B=10^{18}G$ y $B=5\times 10^{18}G$. 
\begin{figure}[h!]
	\centering
	\includegraphics[width=0.49\linewidth]{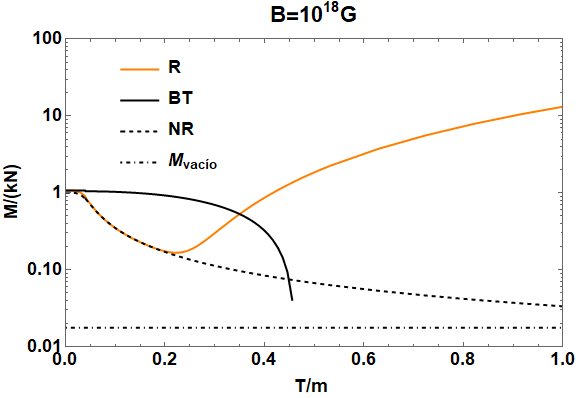}
	\includegraphics[width=0.49\linewidth]{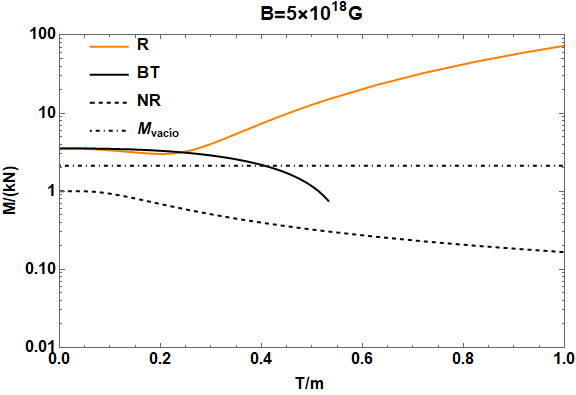}
	\caption{\label{f3.3} Magnetizaci\'on como funci\'on de la temperatura para $N=1.30 \times 10^{39}cm^{-3}$. La l\'inea negra de rayas y puntos corresponde a la contribución del vacío. }
\end{figure}
Para $B=10^{18}G$ la magnetización del vacío es despreciable, y las curvas correspondientes a la magnetización total del gas relativista y los límites NR y BT coinciden para $T=0$ y son iguales a $ k N/\sqrt{1-b}$. A medida que la temperatura aumenta, la magnetización en el l\'imite no relativista decrece y va a cero cuando $T$ tiende a infinito, como vimos en el Cap\'itulo 2. La  magnetizaci\'on del gas relativista muestra este mismo comportamiento hasta $T\sim 0.2\;m$; a partir de este valor $\mathcal{M}^{\pm}(\mu,T,b)$ crece. Este cambio en la monoton\'ia est\'a asociado a las antipart\'iculas, pues alrededor de estas temperaturas su densidad comienza a ser apreciable (v\'ease la Figura \ref{f3.01}), y su aporte a la magnetización tambi\'en. N\'otese que la presencia de las antipart\'iculas aumenta la magnetizaci\'on en varios \'ordenes. 

En el l\'imite de baja temperatura, la magnetizaci\'on tambi\'en decrece cuando $T$ aumenta, pero su comportamiento es bien diferente de los otros dos casos, pues se vuelve negativa alrededor de  $T\sim0.5\;m$ (el punto donde la curva termina). No obstante, esto no implica  que el gas tenga a esas temperaturas un comportamiento diamagnético, sino que es consecuencia de haber despreciado los estados con $s=0$ y $s=-1$, y no es, por tanto, un resultado f\'isicamente correcto. Este comportamiento de la magnetizaci\'on BT refuerza el hecho de que este l\'imite solo es v\'alido para $T\lesssim10^{-3}\;m$.
 
En el caso de  $B=5 \times10^{18}G$, el comportamiento de la magnetizaci\'on es similar, salvo por el hecho de que ahora $\mathcal M_{vac}$ es mayor que el m\'aximo de $\mathcal{M}_{NR}$. En consecuencia, la magnetizaci\'on de los gases relativistas aumenta alrededor de $T=0$. El campo magn\'etico favorece la aproximaci\'on de baja temperatura y vemos que la magnetizaci\'on en este l\'imite coincide en un mayor intervalo de temperatura con la del gas relativista. Esto se debe a que mientras mayor sea el campo magn\'etico m\'as part\'iculas se encuentran en el estado con $s=1$.

Por \'ultimo, es bueno resaltar que los gr\'aficos de la magnetizaci\'on  muestran la importancia de considerar los efectos de las antipart\'iculas y el vac\'io, que normalmente se desprecian. 


Analicemos ahora qu\'e sucede con la magnetizaci\'on del gas relativista cuando $b \rightarrow 0$. Recordemos que en el Cap\'itulo 2 este l\'imite nos condujo al ferromagnetismo de Bose-Einstein. Al hacer $b=0$ en la Ec.(\ref{eq3.7}) se obtiene: 
\begin{equation}\label{eq3.10}
\mathcal{M}_{st}(\mu,T,0)=\sum_{s} \sum_{n=1}^{\infty} \!\frac{z^n+z^{-n}}{n}\bigg\{\frac{m^2ksT}{2\pi^2}K_1(nm/T)+\frac{ksT}{2\pi^2}\int_{m}^{\infty}xK_1(nx/T)\;dx\bigg\},
\end{equation}
\noindent pero al sumar por $s=\pm1$ la expresión anterior se anula. En ausencia de campo magnético, el t\'ermino de la magnetización del vacío también es cero al igual que $\mathcal{M}_{gs}(T,0)=\kappa  N_{gs}(T)$  por encima de la temperatura de condensaci\'on. Sin embargo, para $T<T_c$,  $M_{gs}(T,0) = \kappa N_{gs}(T) \neq 0$ y obtenemos que la magnetizaci\'on total es diferente de cero, aunque $B=0$:
\begin{equation}\label{eq3.11}
\mathcal{M}^{\pm}(\mu,T,0)=\mathcal{M}_{gs}(T,0).
\end{equation} 
Esto evidencia el comportamiento ferromagnético del gas relativista de bosones vectoriales neutros siempre que la condensación esté presente. Al igual que en el l\'imite no relativista, la aparici\'on del ferromagnetismo de Bose-Einstein se asocia a la condensación, ya que en el estado fundamental del gas magnetizado $s=1$.

Para comprobar lo planteado en el párrafo anterior calcularemos como en el Capítulo 2, la capacidad calorífica y la susceptibilidad magnética.
Las expresiones a toda temperatura para la entrop\'ia y la energ\'ia interna por unidad de volumen del gas magnetizado de bosones vectoriales neutros son:
\begin{eqnarray}\label{eq8}
	S^{\pm}(\mu,T,b)=-\frac{2}{T}\Omega_{st}(\mu,T,b)-\frac{\mu}{T}(N^{+}(\mu,T,b)-N^{-}(\mu,T,b))\\\nonumber
	+\sum_{s}\sum_{n=1}^{\infty}\frac{z^n+z^{-n}}{n}\bigg\{\frac{y_0^3}{4\pi^2}[K_1(ny_0/T)+K_3(ny_0/T)]\\
	+\frac{\alpha n}{2\pi^2T}\int_{y_0}^{\infty}\frac{x^3}{\sqrt{x^2+\alpha^2}}K_0(nx/T)dx \bigg\},\nonumber
\end{eqnarray}
\begin{eqnarray}\label{eq9}
	E^{\pm}(\mu,T,b)= -\Omega_{st}(\mu,T,b)+\Omega_{vac}(b)\\\nonumber
	+\sum_{s}\sum_{n=1}^{\infty}\frac{z^n+z^{-n}}{n}\bigg\{\frac{y_0^3T}{4\pi^2}[K_1(ny_0/T)+K_3(ny_0/T)]\\
	+\frac{\alpha n}{2\pi^2}\int_{y_0}^{\infty}\frac{x^3}{\sqrt{x^2+\alpha^2}}K_0(nx/T)dx \bigg\},\nonumber	
\end{eqnarray}
Derivando la Ec.(\ref{eq9}) con respecto a $T$ se obtiene la capacidad calor\'ifica:
\begin{eqnarray}\label{eqCv1}
C_V^{\pm}(\mu,T,b)=S^{\pm}(\mu,T,b)+\sum_{s}\sum_{n=1}^{\infty}\frac{z^n+z^{-n}}{n}\bigg\{\frac{y_0^3}{4\pi^2}[K_1(ny_0/T)+K_3(ny_0/T)]\\\nonumber
+\frac{y_0^4n}{8 \pi^2T}[K_0(ny_0/T)+2K_2(ny_0/T)+K_4(ny_0/T)]+\frac{\alpha n^2}{2\pi^2T^2} \int_{y_0}^{\infty}\frac{x^4K_1(nx/T)dx}{\sqrt{x^2+\alpha^2}}\bigg\} \\\nonumber
-\frac{\mu}{T}\sum_{s}\sum_{n=1}^{\infty}(z^n-z^{-n})\bigg\{\frac{y_0^3}{4\pi^2}[K_1(ny_0/T)+K_3(ny_0/T)]
+\frac{\alpha n}{2\pi^2T}\int_{y_0}^{\infty}\frac{x^3 K_0(nx/T)dx}{\sqrt{x^2+\alpha^2}}\bigg\}.
\end{eqnarray}
\noindent La susceptibilidad magn\'etica se obtiene derivando $\mathcal M^{\pm}$ con respeto a $B$:
\begin{equation}\label{eqChi}
\chi^{\pm}(\mu,T,b)=
\begin{cases}
\chi_{st}(\mu,T,b)+\chi_{vac}(b)+\chi_{gs}(\mu,T,b),& \text{CBE},\\
\chi_{st}(\mu,T,b)+\chi_{vac}(b),& \text{Gas Libre},
\end{cases}
\end{equation}
\noindent donde
\begin{equation}\label{eqChivac}
\chi_{vac}(b)=\frac{k^2m^2}{4\pi}[-b^2+(2b^2-3)\log(1-b)+(2b^2-3)\log(1+b)],
\end{equation} 
\begin{eqnarray}\label{eqChist}
\chi_{st}(\mu,T,b)= \sum_{s}\sum_{n=1}^{\infty}\frac{(z^n-z^{-n})}{n}\bigg\{   \frac{4k^2\alpha^2n}{\pi^2b^2(2-bs)}K_0(ny_0/T)+\\\nonumber \frac{4k^2 s \alpha T(4-3bs)}{\pi^2(2-bs)^3}K_1(ny_0/T)-\frac{3 k^2 s^2 \alpha T}{2\pi^2}\int_{y_0}^{\infty}\frac{x^4K_1(nx/T)dx}{(x^2+\alpha^2)^{5/2}}\bigg\},
\end{eqnarray}
\noindent y
\begin{eqnarray}\label{eqChigs}
\chi_{gs}(\mu,T,b)= \frac{k^2}{m\sqrt{(1-b)^3}}N_{gs}
-\frac{2k^2}{m\sqrt{1-b}}\sum_{s}\sum_{n=1}^{\infty}(z^n-z^{-n})\\\nonumber
\times\bigg\{\frac{m^3s\sqrt{1-bs}}{2\pi^2(2-bs)}K_1(ny_0/T)
+\frac{ms}{4\pi^2}\int_{y_0}^{\infty}\frac{x^4K_1(nx/T)dx}{(x^2+\alpha^2)^{3/2}} \bigg\}.
\end{eqnarray}

Algo a tener en cuenta a la hora de calcular num\'ericamente  la susceptibilidad magn\'etica es que el primer t\'ermino de $\chi_{st}$ diverge para $T<T_c$, sin embargo $\chi^{\pm}(\mu,T,b)$ converge a campo finito debido a que esa divergencia se compensa con el segundo t\'ermino de $\chi_{gs}$. A campo cero,  $\chi_{st}(\mu,T,0)$ diverge para todo $T<T_c$ (los detalles de este an\'alisis pueden ser consultados en el Ap\'endice B). Esto \'ultimo tambi\'en se obtuvo en el Cap\'itulo 2 para la susceptibilidad no relativista, Ec.(\ref{eq48}), y parece estar relacionado con el ferromagnetismo de Bose-Einstein, ya que se observa en las mismas condiciones de temperatura y campo, es decir a $B=0$ y $T<T_c$. 

\begin{figure}[h!]
	\centering
	\includegraphics[width=0.49\linewidth]{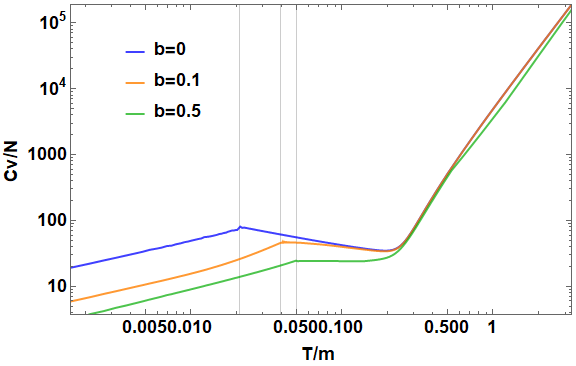}
	\includegraphics[width=0.49\linewidth]{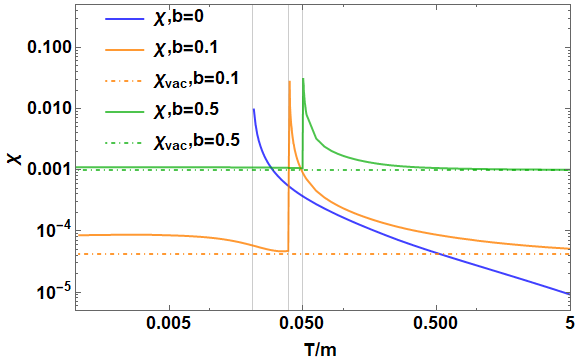}	
	\caption{\label{f3.31} Calor espec\'ifico y susceptibilidad magn\'etica como funci\'on de la temperatura para $N=1.30 \times 10^{39}cm^{-3}$ y varios valores del campo magn\'etico. Las l\'ineas de rayas y puntos corresponden a la contribución del vacío y las líneas continuas verticales a $Tc(b)$.}
\end{figure}

La Figura \ref{f3.31} muestra el calor espec\'ifico y la susceptibilidad magn\'etica como funci\'on de la temperatura para $N=1.30 \times 10^{39}cm^{-3}$ y varios valores del campo magn\'etico. 
Al igual que en el caso no relativista, los picos de $C_{V}^{\pm}(\mu,T,b)$ y $\chi^{\pm}(\mu,T,b)$ ocurren en la temperatura de condensación (las líneas verticales continuas),  indicando el paso del sistema del comportamiento paramagnético al ferromagnético.  Esto refuerza nuestra conclusión de que el comportamiento ferromagnético del gas por debajo de $T_c$ es una consecuencia de la condensación. También podemos ver que la presencia de las antipartículas hace que el calor específico por partícula aumente con la temperatura en vez de tender a $3/2$ como en el caso no relativista. Otro aspecto interesante del panel derecho de la Figura \ref{f3.31} es el comportamiento a altas temperaturas de la susceptibilidad magn\'etica. A $B=0$, $\chi^{\pm}$ disminuye con $T$ hasta hacerse cero como en el caso no relativista. No obstante, a campo magn\'etico finito $\chi^{\pm}\rightarrow\chi_{vac}$ cuando $T$ aumenta. Esto es consistente con el hecho de que la magnetizaci\'on relativista aumenta con la temperatura en lugar de anularse. Como ya vimos, esto es una consecuencia directa de la presencia de una fracci\'on finita de antipart\'iculas en el sistema.

\section{Presiones anisotr\'opicas}

En este epígrafe estudiaremos las ecuaciones de estado del gas magnetizado, en particular el comportamiento con la temperatura de sus presiones paralela y perpendicular:
\begin{equation}\label{eq7}
P^{\pm}_{\parallel}(\mu,T,b)=-\Omega^{\pm}(\mu,T,b),
\end{equation}
\begin{equation}\label{eq7a}
P^{\pm}_{\perp}(\mu,T,b)=P^{\pm}_{\parallel}(\mu,T,b)-\frac{bm}{2k}\mathcal{M}^{\pm }(\mu,T,b).
\end{equation}

La Figura \ref{f3.3} muestra la dependencia de las presiones con la temperatura para $N=1.30 \times 10^{39}cm^{-3}$ y varios valores de campo magn\'etico. A modo de referencia se han dibujado tambi\'en las curvas correspondientes a la presión del vacío. 
\begin{figure}[h!]
	\centering
	\includegraphics[width=0.55\linewidth]{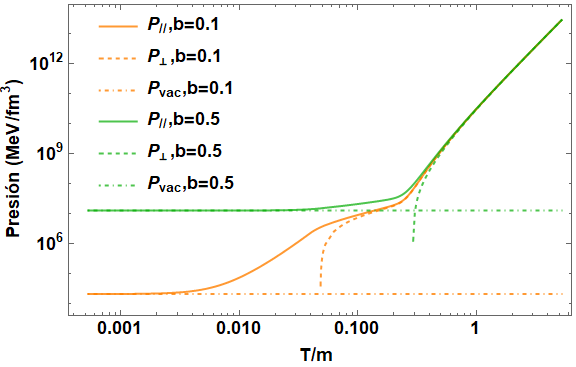}
		\caption{\label{f3.3} Presiones paralela y perpendicular como funciones de la temperatura para $N=1.30 \times 10^{39}cm^{-3}$ y varios valores del campo magn\'etico. Las l\'ineas de rayas y puntos corresponden a la contribución del vacío.  Las l\'ineas continuas constituyen la presión paralela y las discontinuas la presión perpendicular.}
\end{figure}
En este gr\'afico pueden identificarse claramente dos regiones. En la primera, $T>m$, la diferencia entre las presiones paralela y perpendicular es despreciable, as\'i como tambi\'en son despreciables las diferencias que pudieran existir entre las presiones a valores de $B$ distintos. Esta es, por tanto, una regi\'on dominada por la temperatura. 

En la segunda región, $T<m$, es notable la diferencia entre las presiones pues en ella domina el campo magnético.  A medida que la temperatura disminuye $P^{\pm}_{\parallel}$ tiende a la presi\'on del vac\'io, mientras que $P^{\pm}_{\perp}$ se hace negativa. Esto sucede porque a medida que m\'as part\'iculas del gas pasan al condensado, $\Omega^{\pm}_{st}(\mu,T,b) \rightarrow 0$, y $\Omega^{\pm}_{vac}(b)$ y $-B \mathcal M^{\pm}$ se convierten en los t\'erminos dominantes en las presiones paralela y perpendicular respectivamente. Por otra parte, n\'otese en la Figura \ref{f3.3} c\'omo el aumento del campo magn\'etico aumenta la presi\'on del vac\'io y la temperatura a la cual la $P^{\pm}_{\perp}$ se hace cero, facilitando as\'i la ocurrencia del colapso magn\'etico del gas, tal y como sucede tambi\'en en l\'imite no relativista analizado en el Cap\'itulo 2.

\begin{figure}[h!]
	\centering
	\includegraphics[width=0.49\linewidth]{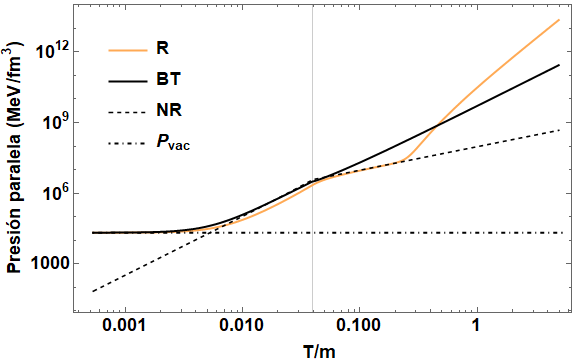}
	\includegraphics[width=0.49\linewidth]{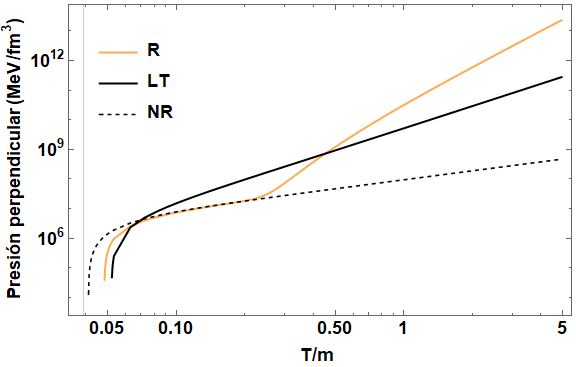}	
	\caption{\label{f3.4} Presiones paralela y perpendicular como funciones de la temperatura para $N=1.30 \times 10^{39}cm^{-3}$ y $B=10^{18}$G. La l\'inea negra de rayas y puntos corresponde a la contribución del vacío.  Las l\'ineas naranja, negra y negra discontinua son las presiones en las diferentes descripciones del gas de bosones vectoriales neutros. La l\'inea horizontal corresponde a la temperatura cr\'itica del CBE.  }
\end{figure}

Las diferencias entre las presiones que resultan del c\'alculo relativista presentado en este cap\'itulo y sus l\'imites NR y BT pueden apreciarse en la Figura \ref{f3.4}. 
Dichas diferencias son, en t\'erminos generales, tres: Primero, la presencia de las antipart\'iculas en la regi\'on de altas temperaturas, que ocasiona una diferencia de varios \'ordenes entre las presiones relativistas y no relativistas. Segundo, que en los dos casos relativistas el comportamiento de la presi\'on paralela a bajas temperaturas es dominado por la presi\'on del vac\'io, mientras que en el l\'imite no relativista $P_{\parallel}$ tiende a cero con $T$. Y tercero, que el valor de la temperatura para la cual $P_{\perp}=0$ se subestima en el l\'imite no relativista, mientras que se sobreestima en la aproximaci\'on de baja temperatura.

    \chapter*{Conclusiones}
\addcontentsline{toc}{chapter}{Conclusiones}

A fin de contribuir al desarrollo de descripciones cada vez más precisas y realistas de los objetos y fenómenos astrofísicos, al iniciar el trabajo que culmina con la presentación de esta tesis nos propusimos estudiar las propiedades termodinámicas de un gas magnetizado de bosones vectoriales neutros a toda temperatura. Como resultado, en la tesis se presentan, por primera vez hasta donde sabemos, las expresiones analíticas exactas para las ecuaciones de estado de un gas magnetizado de bosones vectoriales neutros a toda temperatura, y su límite no relativista, incluidas la magnetización y las segundas derivadas del potencial termodinámico ($C_v$ y $\chi$). Estas EdE están listas para ser utilizadas en cualquier problema (astro)físico que así lo requiera.

El estudio numérico de la termodinámica de este gas en condiciones astrofísicas, permitió ver la influencia relativa de los parámetros del problema (densidad, campo magnético y temperatura) en la ocurrencia de distintos fenómenos. En particular, fueron investigados la condensación de Bose-Einstein, las propiedades magn\'eticas y la anisotrp\'ia en las presiones que es típica de los gases cuánticos magnetizados. A partir de los resultados obtenidos, arribamos a las siguientes conclusiones:

\begin{itemize}

	\item En dependencia de si $T$ es mayor o menor que la masa de los bosones, existen dos reg\'imenes bien diferenciados en el comportamiento del gas. Para $T<<m$, dominan en el sistema los efectos del campo magn\'etico, mientras que si $T>>m$ dominan los efectos de la temperatura, siendo el m\'as importante de ellos la existencia de una fracci\'on no despreciable de antipart\'iculas en el caso del gas relativista.
	
	\item En los dos casos estudiados, la transición de fase al CBE depende de la temperatura, la densidad de partículas y el campo magnético, de manera tal que al aumentar la densidad de partículas o el campo magnético, así como al disminuir la temperatura, el condensado puede aparecer. 
	
	\item La relevancia de las antipart\'iculas en el gas relativista se evidencia claramente en todas las magnitudes. En general, la densidad de antipart\'iculas deja de ser despreciable alrededor de $T\sim0.2\;m$ para $N=1.30\times10^{39}$cm$^{-3}$, aunque sus efectos se manifiestan con más fuerza para $T \gtrsim m$, siendo ambos valores de temperatura, al parecer, bastante independientes del campo magn\'etico.

	Esto \'ultimo cobra especial importancia en los casos de la magnetizaci\'on y la presi\'on del gas, pues la aparici\'on de las antipart\'iculas en el r\'egimen de altas temperaturas implica un aumento de varios \'ordenes en ambas magnitudes. De manera que las ant\'iparticulas podr\'ian estar conectadas tanto a la producci\'on y mantenimiento de los altos campos magn\'eticos estelares, como a la estabilidad gravitacional de los objetos astron\'omicos.
	
	\item Por debajo de la temperatura crítica del CBE, el gas muestra, para todas las descripciones estudiadas, una magnetización espontánea. La aparici\'on del ferromagnetismo de Bose-Einstein  se asocia a la condesaci\'on, lo que fue demostrado a trav\'es del estudio de $C_v$ y $\chi$. La capacidad de los gases vectoriales de magnetizarse espont\'aneamente podr\'ia contarse tambi\'en entre las causas de que originan y mantienen el campo magn\'etico en entornos astrof\'isicos.
	
	\item Como sucede con otros gases cuánticos magnetizados, la presencia del campo magnético impone su simetría axial al sistema y separa las presiones en dos componentes, una a paralela y otra perpendicular al eje magnético. La presión paralela aumenta con el campo magnético, mientras que la perpendicular disminuye. Bajo ciertas condiciones, la presión perpendicular puede ser negativa y el sistema se vuelve inestable. Como se mostró, esta inestabilidad es causada por el campo magnético, mientras que la temperatura se opone a ella. Por otro lado, aumentar la densidad de partículas también desestabiliza el gas. Esto impone un límite superior a las densidades de bosones permitidas en los objetos astronómicos para una temperatura y un campo magnético dado.
	
\end{itemize}

Finalmente, la comparación entre el cálculo relativista general y los límites no relativista y de baja temperatura permitió establecer los rangos de validez de dichas aproximaciones y sus efectos en el sistema estudiado. Aunque las expresiones obtenidas para las magnitudes termodin\'amicas en el l\'imite no relativista son v\'alidas para toda temperatura, esta aproximaci\'on no toma en cuenta la producci\'on de antipart\'iculas a altas temperaturas, ni los efectos de vac\'io, que pueden llegar a ser muy importantes para el gas magnetizado en la región de bajas temperaturas. En cambio, encontramos que el l\'imite de baja temperatura solo es aplicable para campos magnéticos fuertes. Las limitantes halladas en ambas aproximaciones con respecto al c\'alculo relativista a toda temperatura demuestran su importancia, y reafirman la necesidad y relevancia del trabajo llevado a cabo en la presente tesis.

\chapter*{Recomendaciones}
\addcontentsline{toc}{chapter}{Recomendaciones}

 Al contar con las ecuaciones de estado para el gas de bosones vectoriales neutros a toda temperatura, podemos darle continuidad a los estudios previos de las estrellas de bosones y obtener las modificaciones que sufren los observables a consecuencia de la temperatura, así como investigar  otros fenómenos relacionados con esta, como el enfriamiento de la estrella y diferentes tipos de emisión.

    \appendix
    \addcontentsline{toc}{chapter}{Apéndices}
     \chapter{C\'alculo del calor espec\'ifico  no relativista}
\label{appA}

El procedimiento para calcular el calor espec\'ifico depende de si estamos en presencia de la fase condensada o no. Comencemos con el caso en que ya hay condensado, ya que el cálculo es más simple. En en el condesado $C_{V_{NR}}$ es: 
\begin{equation}\label{eq36}
C_{V_{NR}}=\frac{\partial}{\partial T}\bigg[-\frac{3}{2}\Omega_{NR}(\mu,T,B)-kB[N_{+}-N_{-}]\bigg],
\end{equation}
\noindent luego de derivar:
\begin{eqnarray}\label{eq37}\nonumber 
C_{V_{NR}}(\mu,T,B)=\frac{3}{2}S_{NR}(\mu,T,B)-\frac{3}{2}\frac{kB}{T}[N_{+}-N_{-}]+\frac{kB}{T^2}\left(\frac{mT}{2\pi}\right)^{3/2}\\
\times[(\mu+kB)g_{1/2}(z_{+})-(\mu-kB)g_{1/2}(z_{-})],
\end{eqnarray}
\noindent sustituyendo $\mu=-kB$  queda:
\begin{eqnarray} \label{eq38} 
C_{V_{NR}}(-kB,T,B)=\frac{3}{2}S_{NR}(-kB,T,B)-\frac{3}{2}\frac{kB}{T}[N_{+}-N_{-}]+2\left(\frac{kB}{T}\right)^2\left(\frac{mT}{2\pi}\right)^{3/2}g_{1/2}(z_{-}).
\end{eqnarray}
\noindent En la fase no condensada $N=N_{+}+N_{-}+N_{0}$ ya que $N_{gs}=0$  y se cumple:
\begin{eqnarray}\label{eq39}
\bigg(\frac{mT}{2\pi}\bigg)^{3/2}=\frac{1}{g_{3/2}(z_{+})+g_{3/2}(z_{-})+g_{3/2}(z_{0})}.
\end{eqnarray}
Utilizando la identidad anterior podemos reescribir la densidad de energía Ec.(\ref{eq2.12}):
\begin{eqnarray}\label{eq39.1}
\frac{E_{NR}(\mu,T,B)}{N}=\frac{3}{2}T\bigg[\frac{g_{5/2}(z_{-})+g_{5/2}(z)+g_{5/2}(z_{+})}{g_{3/2}(z_{-})+g_{3/2}(z)+g_{3/2}(z_{+})}\bigg]-k B \bigg[\frac{g_{3/2}(z_{-})-g_{3/2}(z_{+})}{g_{3/2}(z_{-})+g_{3/2}(z)+g_{3/2}(z_{+})} \bigg],
\end{eqnarray}
\noindent y escribir el calor específico como:
\begin{eqnarray}\label{eq40}\nonumber
\frac{C_{V_{NR}}}{N}= \frac{3}{2}\bigg[\frac{g_{5/2}(z_{-})+g_{5/2}(z)+g_{5/2}(z_{+})}{g_{3/2}(z_{-})+g_{3/2}(z)+g_{3/2}(z_{+})}\bigg]+ \frac{3}{2} T \frac{\partial}{\partial T} \bigg[\frac{g_{5/2}(z_{-})+g_{5/2}(z)+g_{5/2}(z_{+})}{g_{3/2}(z_{-})+g_{3/2}(z)+g_{3/2}(z_{+})}\bigg]- \\
-k B \frac{\partial}{\partial T} \bigg[\frac{g_{3/2}(z_{-})-g_{3/2}(z_{+})}{g_{3/2}(z_{-})+g_{3/2}(z)+g_{3/2}(z_{+})} \bigg].
\end{eqnarray}
\noindent Vamos a llamarle $ D_{1}$ a la  derivada del segundo término de (\ref{eq40})  y   $ D_{2}$ a la del tercer término, ya que las calcularemos por separado. Como, $ \frac {\partial} {\partial T} [g_{3/2} (z_{-}) + g_{3/2} (z) + g_{3/2} (z_{+})] =-\frac{3}{2}\frac{1}{T} \left(\frac{2\pi}{mT}\right)^{3/2}$ y usando (\ref{eq38}) obtenemos:
\begin{eqnarray}\label{eq41} \nonumber
\frac{\partial}{\partial T}[g_{3/2}(z_{-})]=-\frac{3}{2}\frac{1}{T}g_{3/2}(z_{-}),\\ \nonumber
\frac{\partial}{\partial T}[g_{3/2}(z)]=-\frac{3}{2}\frac{1}{T}g_{3/2}(z),\\
\frac{\partial}{\partial T}[g_{3/2}(z_{+})]=-\frac{3}{2}\frac{1}{T}g_{3/2}(z_{+}).
\end{eqnarray}
\noindent Utilizando la siguiente fórmula de recurrencia de la función polilogartímica $ z_{i} \frac {\partial} {\partial z_{i}} [g_{\nu} (z_{i})] = g_{\nu-1} (z_{i}) $ se obtiene:
\begin{eqnarray}\label{eq42}
\frac{\partial}{\partial T}g_{3/2}(z_{i})=\frac{\partial }{\partial z_{i}}[g_{3/2}(z_{i})] \frac{\partial z_{i}}{\partial T}
\Longrightarrow \frac{1}{z_{i}}\frac{\partial z_{i}}{\partial T} = \frac{\partial}{\partial T} [g_{3/2}(z_{i})] \frac{1}{g_{1/2}(z_{i})},
\end{eqnarray}
luego:
\begin{eqnarray}\label{eq43} \nonumber
D_{1}= \frac{ \frac{\partial}{\partial z_{-}}[g_{5/2}(z_{-})] \frac{\partial z_{-}}{\partial T}+ \frac{\partial}{\partial z}[g_{5/2}(z)] \frac{\partial z}{\partial T}+ \frac{\partial}{\partial z_{+}}[g_{5/2}(z_{+})] \frac{\partial z_{+}}{\partial T}}{g_{3/2}(z_{-})+g_{3/2}(z)+g_{3/2}(z_{+})} \\
- \frac{g_{5/2}(z_{-})+g_{5/2}(z)+g_{5/2}(z_{+})}{[g_{3/2}(z_{-})+g_{3/2}(z)+g_{3/2}(z_{+})]^{2}}\frac{\partial}{\partial T} [g_{3/2}(z_{-})+g_{3/2}(z)+g_{3/2}(z_{+})].
\end{eqnarray}
\noindent Haciendo uso de $\frac{\partial}{\partial T} [g_{\nu}(z_{\pm})]=-\frac{(\mu\pm k B)}{T^{2}}g_{\nu-1}(z_{\pm}),\;\frac{\partial}{\partial T}[g_{\nu}(z)]=-\frac{\mu}{T^{2}}g_{\nu-1}(z),$  (\ref{eq41}) y (\ref{eq42}) es f\'acil llegar a:
\begin{eqnarray}\label{eq44}
D_{1}= -\frac{3}{2}\frac{1}{T}\frac{\frac{[g_{3/2}(z_{-})]^{2}}{g_{1/2}(z_{-})}+\frac{[g_{3/2}(z)]^{2}}{g_{1/2}(z)}+\frac{[g_{3/2}(z_{+})]^{2}} {g_{1/2}(z_{+})}}{g_{3/2}(z_{+})+g_{3/2}(z)+g_{3/2}(z_{+})}+ \frac{3}{2}\frac{1}{T} \frac{g_{5/2}(z_{-})+g_{5/2}(z)+g_{5/2}(z_{+})}{g_{3/2}(z_{-})+g_{3/2}(z)+g_{3/2}(z_{+})},
\end{eqnarray}
\begin{eqnarray}\label{eq45}
D_{2}=0.
\end{eqnarray}
\noindent Finalmente obtenemos el calor espec\'ifico en la fase no condensada:
\begin{eqnarray}\label{eq46}
\frac{C_{V_{NR}}(\mu,T,B)}{N}=\frac{15}{4}\frac{g_{5/2}(z_{-})+g_{5/2}(z)+g_{5/2}(z_{+})}{g_{3/2}(z_{-})+g_{3/2}(z)+g_{3/2}(z_{+})}-\frac{9}{4}
\frac{\frac{[g_{3/2}(z_{-})]^{2}}{g_{1/2}(z_{-})}+\frac{[g_{3/2}(z)]^{2}}{g_{1/2}(z)}+\frac{[g_{3/2}(z_{+})]^{2}}
	{g_{1/2}(z_{+})}}{g_{3/2}(z_{+})+g_{3/2}(z)+g_{3/2}(z_{+})}.
\end{eqnarray}

     \chapter{An\'alisis de convergencia de la susceptibilidad magn\'etica relativista en presencia del condensado}
\label{appB}

Analicemos la convergencia del primer t\'ermino de $\chi_{st}$ en la Ec. (\ref{eqChist}):
\begin{equation}\label{Eqb1}
\sum_{s}\sum_{n=1}^{\infty}   \frac{k^2m^2s^2}{\pi^2(2-bs)^2}(z^n-z^{-n})K_0(nm\beta\sqrt{1-bs}).
\end{equation}
\noindent  Sumando por $s=-1,0,1$ queda:
\begin{equation}
\sum_{n=1}^{\infty}\bigg\{ \frac{k^2m^2}{\pi^2(2-b)^2}(z^n-z^{-n})K_0(nm\beta\sqrt{1-b})+\frac{k^2m^2}{\pi^2(2+b)^2}(z^n-z^{-n})K_0(nm\beta\sqrt{1+b})\bigg\},
\end{equation}
\noindent cuando $n\rightarrow\infty$,$\;$ $K_0(nm\beta\sqrt{1\pm b})\sim \frac{\sqrt{\pi}e^{-nm\beta\sqrt{1\pm b}}}{\sqrt{2nm\beta\sqrt{1\pm b}}}$; por tanto, para $n>>1$ se cumple que:
\begin{equation}
a_n(s=\pm1)=(e^{n\mu\beta}-e^{-n\mu\beta})K_0(nm\beta\sqrt{1\pm b})\approx\frac{\sqrt{\pi}(e^{n\beta(\mu-m\sqrt{1\pm b})}+e^{-n\beta(\mu+m\sqrt{1\pm b})})}{\sqrt{2nm\beta\sqrt{1\pm b}}}.
\end{equation}
\noindent En el condensado $\mu=m\sqrt{1-b}$, por tanto:
\begin{equation}\label{Eqb4}
a_n(s=-1)=\frac{\sqrt{\pi}(e^{nm\beta(\sqrt{1-b}-\sqrt{1+ b})}+e^{-nm\beta(\sqrt{1-b}+\sqrt{1+ b})})}{\sqrt{2nm\beta\sqrt{1+b}}},
\end{equation}
\noindent y 
\begin{equation}\label{Eqb5}
a_n(s=1)=\frac{\sqrt{\pi}(1+e^{-nm\beta2\sqrt{1-b}})}{\sqrt{2nm\beta\sqrt{1-b}}}.
\end{equation}
\noindent Cuando $n\rightarrow\infty$, la Ec.(\ref{Eqb4}) tiende a cero exponencialmente, mientras que la Ec.(\ref{Eqb5}) tiende a cero como $1/\sqrt{n}$, por tanto, la suma infinita de $a_n(s=1)$ no converge. En los otros t\'erminos de $\chi_{st}$ con $s=1$ pasa lo mismo, pero como est\'an divididos por $n$, tienden a cero como $\frac{1}{n^{3/2}}$  y no hay problemas de convergencia. Luego $\chi_{st}$ no converge para $T<T_c$. Para campo finito la convergencia de $\chi^{\pm}$ se logra porque el t\'ermino infinito en $\chi_{st}$  se compensa con el segundo sumando  de $\chi_{gs}$:
\begin{equation}\label{Eqb6}
-\sum_{s}\sum_{n=1}^{\infty}   \frac{k^2m^2s\sqrt{1-bs}}{\pi^2\sqrt{1-b}(2-bs)}(z^n-z^{-n})K_0(nm\beta\sqrt{1-bs}),
\end{equation}
\noindent que tambi\'en tiende a cero como $1/\sqrt{n}$ cuando $n\rightarrow\infty$. Sin embargo, 
para $b=0$ la Ec.(\ref{Eqb6}) se anula cuando sumamos por $s$, no hay nada que compense el infinito de  $\chi_{st}$ y, como consecuencia, $\chi^{\pm}(\mu,T,0)$ diverge para $T<T_c$.

%
%
%
%


  \backmatter

  \bibliographystyle{ieeetr}  
  \bibliography{bib} 	

\end{document}